\documentclass[%
 reprint,
 amsmath,amssymb,
 aps,
]{revtex4-2}

\usepackage{dcolumn}
\usepackage{bm}
\usepackage{hyperref}

\usepackage{amsmath,amssymb,amsfonts, bm, amsthm}
\usepackage{tensor,cancel}
\usepackage[Symbolsmallscale]{upgreek}
\usepackage{alphabeta}
\usepackage{lipsum}

\usepackage{physics,siunitx}

\usepackage[dvipsnames]{xcolor}

\usepackage{hyperref}
    \hypersetup{
        colorlinks=true,
        linkcolor=blue,
        citecolor=blue,
        filecolor=Orange,      
        urlcolor=red,
        pdftitle={Overleaf Example},
        pdfpagemode=FullScreen,
        }

\usepackage{graphicx}
    \graphicspath{{figures/}}
 
 \usepackage{tikz,pgfplots,tikz-3dplot}
 	\pgfplotsset{compat=1.15}
    \usetikzlibrary{arrows,decorations,backgrounds,petri,3d}
    \usetikzlibrary{shapes.arrows}
	\usetikzlibrary{shapes.geometric, shapes.arrows, positioning, math}
	\tikzset{MyArrow/.style={single arrow, draw, minimum width=5ex, minimum height=10ex, inner sep=1ex, text height=1ex, text depth=0ex, single arrow head extend=1ex} }
    
\usepackage{comment}

\newcommand{\ensuremathcommand}[2]{\let#2#1\renewcommand{#1}{\ensuremath{#2}}}

\newcommand{\txt}[1]{_\text{#1}}

\newcommand{\half}{\ensuremath{\tfrac{1}{2}}}

\newcommand{\lrp}[1]{\ensuremath{\left( #1 \right)}}
\newcommand{\lrb}[1]{\ensuremath{\left[ #1 \right]}}
\newcommand{\lrB}[1]{\ensuremath{\left\{ #1 \right\}}}

\renewcommand{\c}{\cdot}
\renewcommand{\d}{\mathrm{d}}

\newcommand{\f}{\frac}

\newcommand{\m}[1]{\overline{#1}}
\newcommand{\n}{\widetilde}

\newcommand{\q}{\quad}
\renewcommand{\qq}{\qquad}

\renewcommand{\t}{\times}

\renewcommand{\L}{\mathcal{L}}
\newcommand{\M}{\mathcal{M}}
	\newcommand{\MN}{\mathcal{MN}}
\newcommand{\N}{\mathcal{N}}

\renewcommand{\P}{\mathcal{P}}

\renewcommand{\S}{\mathcal{S}}
\newcommand{\T}{\mathrm{T}}

\newcommand{\W}{\mathcal{W}}

\usepackage{afterpage}
\usepackage[skip=5pt, indent=15pt]{parskip}

\begin{document}

\preprint{APS/123-QED}

\title{Cosmological Parameter Estimation with \\ Sequential Linear Simulation-based Inference}%

\author{N.\;G.\;Mediato\hspace{1.3pt}-Diaz}
\affiliation{%
 Cavendish Laboratory, University of Cambridge,\\ JJ Thomson Avenue, Cambridge, CB3 0HE
}%

\author{W.\;J.\;Handley}
\affiliation{
 Kavli Institute for Cosmology, University of Cambridge,\\ Madingley Road, Cambridge, CB3 0HA
}%

\begin{abstract}
\vspace{2mm}
We develop the framework of Linear Simulation-based Inference (LSBI), an application of simulation-based inference where the likelihood is approximated by a Gaussian linear function of its parameters. We obtain analytical expressions for the posterior distributions of hyper-parameters of the linear likelihood in terms of samples drawn from a simulator, for both uniform and conjugate priors.  This method is applied sequentially to several toy-models and tested on emulated datasets for the Cosmic Microwave Background temperature power spectrum. We find that convergence is achieved after four or five rounds of $\mathcal{O}(10^4)$ simulations, which is competitive with state-of-the-art neural density estimation methods. Therefore, we demonstrate that it is possible to obtain significant information gain and generate posteriors that agree with the underlying parameters while maintaining explainability and intellectual oversight.

\vspace{2mm}
\begin{description}
\item[Key words]
 cosmology, Bayesian data analysis, simulation-based inference, CMB power spectrum
\end{description}
\end{abstract}

\maketitle


\section{Introduction }\label{sec:introduction}

In many astrophysical applications, statistical models can be simulated forward, but their likelihood functions are too complex to calculate directly. Simulation-based inference (SBI)   \cite{cranmer2020frontier} provides an alternative way to perform Bayesian analysis on these models, relying solely on forward simulations rather than likelihood estimates. However, modern cosmological models are typically expensive to simulate and datasets are often high-dimensional, so traditional methods like the Approximate Bayesian Computation (ABC) \cite{rubin1984bayesianly}, which scale poorly with dimensionality, are no longer suitable for parameter estimation. Improvements to ABC, such as the inclusion of Markov-chain Monte Carlo \cite{marjoram2003markov} and Sequential Monte Carlo \cite{sisson2007sequential} methods, can also have limited effectiveness for large datasets. 

In the past decade, machine learning techniques have revolutionized the field of SBI \cite{cranmer2020frontier}, enabling a reduction in the number of simulator calls required to achieve high-fidelity inference, and providing an efficient framework to analyze complex data and expensive simulators. In particular, we highlight Density Estimation Likelihood-free Inference (DELFI) \cite{papamakarios2016fast,alsing2019fast} and Truncated Marginal Neural Ratio Estimation (TMNRE) \cite{cole2022fast}. Given a set of simulated parameter-data pairs, these algorithms learn a parametric model for the joint and the likelihood-to-evidence ratio respectively, via neural density estimation (NRE) techniques \cite{lemos2023robust,papamakarios2019neural}. Furthermore, recent applications of SBI to high-dimensional cosmological and astrophysical datasets
\cite{dupourque2023investigating, gatti2024dark, crisostomi2023neural, christy2024applying, harnois2024kids, moser2024simulation, novaes2024cosmology, fischbacher2024texttt}
demonstrate that these algorithms are rapidly establishing themselves as a standard machine learning technique in the field.

However, the use of neural networks presents some disadvantages, the most significant of which is their lack of {\it explainability}. This means that most neural networks are treated as a `black box', where the decisions taken by the artificial intelligence in arriving at the optimized solution are not known to researchers, which can hinder intellectual oversight \cite{castelvecchi2016can}. This problem affects the algorithms discussed above, as NRE constitutes an unsupervised learning task, where the artificial intelligence is given unlabeled input data and allowed to discover patterns in its distribution without guidance. This combines with the problem of over-fitting, where the neural network may attempt to maximize the likelihood without regard for the physical sensibility of the output. Current algorithms for simulation-based inference can produce overconfident posterior approximations \cite{hermans2021trust}, making them possibly unreliable for scientific use. 

The question naturally arises as to whether it is possible to achieve fast and credible simulation-based inference without dependence on neural networks. Such a methodology might allow researchers to acquire control over the inference process and better assess whether their approximations produce overconfident credible regions. Moreover, disentangling ML from SBI can be pedagogically useful in explaining SBI to new audiences by separating its general principles from the details of neural networks.

This paper takes a first step in this direction. In Section \ref{sec:theory} we develop the theoretical framework of Linear Simulation-based Inference (LSBI), an application of likelihood-free inference where the model is approximated by a linear function of its parameters and the noise is assumed to be Gaussian with zero mean. In Section \ref{sec:toy_models}, several toy models are investigated using LSBI, and in Section \ref{sec:results} the method is applied to parameter estimation for the Cosmic Microwave Background (CMB) power spectrum.

\section{Theory }\label{sec:theory}

\subsection{The Linear Approximation }\label{sec:linear_model}

Let us consider a $d$-dimensional dataset $D$ described by a model $\M$ with $n$ parameters $\theta = \{\theta_i\}$. We assume a Gaussian prior $\theta|\M \sim \N(\mu,\Sigma )$ with known mean and covariance.  The likelihood $\L_D(\theta)$ for an arbitrary model is generally intractable, but in this paper, we approximate it as a homoscedastic Gaussian (thus neglecting any parameter-dependence of the covariance matrix),
\begin{equation}\label{eq:likelihood1}
	D|\theta,\M \sim \N(\M(\theta),C ).
\end{equation} 
Furthermore, we approximate the model linearly about a fiducial point $\theta_0$, such that $\M(\theta) \approx m + M\theta$, where $M \equiv \grad_{\theta}\,\M\,\big\vert_{\theta_0}$ and $m \equiv \M(\theta_0) - M\theta_0$. Under these assumptions, the resulting likelihood is
\begin{equation} \label{eq:likelihood}
	D|\theta,\M \sim \N(m + M\theta,C )
\end{equation}
If we knew the likelihood hyper-parameters $m$, $M$, and $C$, the posterior distribution could be found and would also be Gaussian
\begin{equation}\label{eq:posterior}
	\theta| D,\M\sim \mathcal{N}(\mu_\mathcal{P},\Sigma_\mathcal{P})
\end{equation}
where 
\begin{align}
	\label{posterior_cov} &	\Sigma_\mathcal{P}^{-1} \equiv M^\T C^{-1} M + \Sigma^{-1} ,				\\[1mm]
	\label{posterior_mean}&	\mu_\mathcal{P} \equiv \mu +  \Sigma_\mathcal{P}	 M^\T C^{-1} (D-m-M\mu) .\
\end{align}
Similarly, the evidence would given by
\begin{equation} \label{eq:evidence}
	D|\M \sim \mathcal{N}(m+M\mu, C+M\Sigma M^\T).
\end{equation}
Nevertheless, $m$, $M$, and $C$ are unknown, so we must obtain their distributions before computing the posterior.\\

\subsection{Linear Simulation-based Inference} \label{sec:LSBI}

The goal of SBI is to obtain an approximate form of the likelihood through the simulator itself. For the purposes of this paper, the simulator $\S_\M$ is understood to be a stochastic function that takes the model parameters as input and yields a set of noisy simulated data
\begin{equation} \label{eq:simulator}
	\theta \q \overset{\S_\M}{\xrightarrow{\hspace*{1.5cm}}} \q D\txt{sim}.
\end{equation}
Hence, upon fixing a set of parameter samples $\{\theta^{(i)}\}$, we can run the simulator on each to obtain a set of simulated data samples $\{D^{(i)}\txt{sim}\}$. Crucially, these are distributed as the true likelihood $D^{(i)}\txt{sim}\sim \mathcal{L}(\,\cdot\,|\theta)$. Thus, one may obtain a numerical estimation of the likelihood through the simulator by learning the probability density of the pairs $\{\theta^{(i)},D\txt{sim}^{(i)}\}$ for a sufficient number of simulator runs. Here, we follow this strategy, except that the assumption of linearity in Eq.~\ref{eq:likelihood} avoids the need for machine-learning tools. This linear analysis applied to SBI is not available in the literature, although some recent works are similar, such as SELFI \cite{leclercq2019primordial} or MOPED \cite{heavens2000massive}.

We first draw $k$ parameter vectors $\{\theta^{(i)}\}$;  since we are estimating the likelihood, we may draw these from an arbitrary distribution that does not need to be the model prior $\pi(\theta)$. Then, for each $\theta^{(i)}$ the simulator is run to produce the corresponding data vector $D^{(i)}$. We define the first- and second-order statistics
\begin{align}\label{eq:means}
	\m{\theta} = \frac{1}{k}\sum^k_{i=1} \theta^{(i)}, \qq \m{D} = \frac{1}{k}\sum^k_{i=1} D^{(i)},
\end{align}
and
\begin{align}
	\Theta 	&= \frac{1}{k}\sum^k_{i=1} (\theta^{(i)}-\m{\theta})(\theta^{(i)}-\m{\theta})^\T, \\
    \Delta 	&= \frac{1}{k}\sum^k_{i=1} ( D^{(i)}-\m{ D})( D^{(i)}-\m{D})^\T, \\
    \Psi 	&= \frac{1}{k}\sum^k_{i=1} ( D^{(i)}-\m{D})(\theta^{(i)}-\m{\theta})^\T, \label{eq:covs}
\end{align}
Then, by expanding the joint likelihood,
\begin{align}
\begin{aligned}
	p(\{D^{(i)}\}|\{\theta^{(i)}&\},m,M,C)\\
	& \equiv {\textstyle\prod}_{i=1}^k  p(D^{(i)}|\theta^{(i)},m,M,C),	
\end{aligned}
\end{align}
From this result, and a choice of broad uniform priors for $m$, $M$ and $C$, we find the distributions
\begin{align}
	\label{eq:p_m}
	m|M,C,S &\sim \N\lrp{\m D-M\m\theta,\tfrac{C}{k} }\\[2mm]
	\label{eq:p_M}
	M|C,S &\sim \MN\lrp{\Psi\Theta^{-1},\tfrac{C}{k},\Theta^{-1} }\\[2mm]
	\label{eq:p_C}
	C|S &\sim \W^{-1}\lrp{k(\Delta -  \Psi \Theta^{-1}\Psi^\T), \nu }
\end{align}
where $S = \{(\theta^{(i)},D^{(i)})\}$ are the simulated parameter-data pairs, and $\nu = k-d-n-2$. $\mathcal{MN}(M,U,V)$ stands for the matrix normal distribution with mean $M$ and covariance $U\otimes V$, and $\mathcal{W}^{-1}(\Lambda,\nu)$ stands for the inverse Wishart distribution with scale matrix $\Lambda$ and $\nu$ degrees of freedom \cite{gupta2018matrix}. Note that $\nu>d-1$, so there is a minimum number of simulations $k\txt{min} = n+2d+2$ required to yield well-defined distributions. Appendix \ref{ap:A} gives the details of the equivalent result for a choice of conjugate priors for $m$, $M$, and $C$,
\begin{align}
	 m|M,C,\{\theta^{(i)}\}  &\sim  \N(0,C),  \\[2mm]
	 M|C,\{\theta^{(i)}\}  &\sim  \MN(0,C,\Theta^{-1}),  \\[2mm]
	 C|\{\theta^{(i)}\}  &\sim  \W^{-1}(C_0,\nu_0).
\end{align}

The posterior can finally be estimated
\begin{align}\label{eq:posterior_est}
	\P_D(\theta) &= \int \P_D(\theta|m,M,C)\times p(m,M,C) \ \d m\, \d M\, \d C \notag\\
	&\approx \Big\langle \P_D(\theta|m,M,C)\Big\rangle_{m,M,C}.
\end{align}
The average can be computed by drawing $N$ exact samples $(m^{(I)},M^{(I)},C^{(I)})$ from Eqs.~\ref{eq:p_m} -- \ref{eq:p_C}, where $N$ is large enough to guarantee convergence. For $N>1$, the resulting posterior is a Gaussian mixture of $N$ components. Since each sample is independent, this calculation can be made significantly faster by parallelization, allowing a large $N$ without much effect on the computational efficiency of the calculation.

\subsection{Sequential LSBI}\label{sec:seqLSBI}

As discussed in Section \ref{sec:linear_model}, the linear approximation is only applicable within a localized region surrounding the fiducial point $\theta_0$. Given that the prior distribution is typically broad, this condition is not often satisfied. Consequently, when dealing with non-linear models, LSBI will truncate the model to its linear expansion, thereby computing the posterior distribution corresponding to an ‘effective’ linear likelihood function. This truncation results in a less constraining model, leading to a broader ‘effective’ posterior distribution relative to the ‘true’ posterior.

However, since the simulation parameters $\{\theta^{(i)}\}$ can be drawn from any non-singular distribution, independent of the prior, LSBI can be performed on a set of samples generated by simulations that are proximal to the observed data, i.e., a narrow distribution with $\theta_0$ near the true parameter values. A natural choice for this distribution is the `effective' LSBI posterior. This leads to the concept of Sequential LSBI, wherein LSBI is iteratively applied to simulation samples drawn from the posterior distribution of the previous iteration, with the initial iteration corresponding to the prior distribution.

It is worth noting that this method suffers from two disadvantages compared to plain LSBI. Firstly, the algorithm is no longer amortized, as subsequent iterations depend on the choice of $D\txt{obs}$. Secondly, as the sampling distribution becomes narrower, $\Theta$ becomes smaller, resulting in a broader distribution for $M$. Thus, the value of $N$ may need to be increased accordingly.

The evidence may be evaluated similarly. Thus, if the procedure is repeated for a different model $\M'$, the Bayes' ratio between the two models may be calculated,
\begin{equation}
	\mathcal{B} = \f{\langle p(D\txt{obs}|\M)\rangle_{m,M,C}}{\langle p(D\txt{obs}|\M')\rangle_{m',M',C'}}
\end{equation}
Nevertheless, this calculation is inefficient for large datasets, so a data compression algorithm is proposed in Appendix \ref{ap:C}, although it is not investigated further in this paper.

\section{Results}\label{sec:toy_models}

In the development of LSBI, we have made two foundational assumptions about the nature of the underlying likelihood and model:
\begin{itemize}
	\item the model $\M(\theta)$ is approximated as a linear function of the parameters.
	\item the likelihood $\L_D(\theta)$ can be accurately approximated by a homoscedastic Gaussian distribution (Eq.~\ref{eq:likelihood1}),
\end{itemize}
In this section, we test the resilience of LSBI against deviations from these assumptions by applying the procedure to several toy models, as well as the CMB temperature power spectrum. These toy models were implemented with the help of the Python package \texttt{lsbi}, currently under development by W.J.~Handley, and tools from the \texttt{scipy} library. To simulate the cosmological power spectrum data, we use the \texttt{cmb\_tt} neural emulator from CosmoPowerJAX \cite{piras2023cosmopower,spurio2022cosmopower}. 

In addition to LSBI, the parameter posteriors are also calculated via nested sampling with \texttt{dynesty} \cite{speagle2020dynesty,koposov2022joshspeagle,higson2019dynamic} for comparison. The plots were made with the software \texttt{getdist} \cite{lewis2019getdist}.  Unless otherwise stated, the calculations in this section use broad uniform priors.

\subsection{Toy Models}

For simplicity, the prior on the parameters is a standard normal $\theta\sim \N(0,I_n)$, and the samples for the simulations are taken directly from this prior. To generate the model, we draw the entries of $m$ from a standard normal, whereas the entries of $M$ have mean 0 and standard deviation $1/d$. The covariance $C$ is drawn from $\W^{-1}(\sigma^2 I_d, d+2)$, where $\sigma=0.5$. The number of samples $N$ taken from posterior distributions of $m$, $M$, and $C$ depends on the dataset's dimensionality $d$; generally, we choose the highest value allowing for manageable computation time.

Our starting point is a 50-dimensional dataset with a quadratic model of $n=4$ parameters,
\begin{equation}\label{eq:model}
	\M(\theta) = m + M\theta + \theta^\T Q\, \theta
\end{equation}
and Gaussian likelihood, where $m$ and $M$ are as above, and $Q$ is a $n \times d \times n$ matrix with entries drawn form $\N(0,1/d)$. The noise is now Gaussian with covariance $C$. At each round, we perform LSBI with $k=2500$ simulations to obtain an estimate for the posterior distribution, where the sampling distribution of the parameter sets $\{\theta^{(i)}\}$ is the posterior of the previous round. The posterior is calculated for a set of `observed' data, which are determined by applying the model and noise to a set of `real' parameters $\theta^*$. We also calculate the KL divergence ($\mathcal{D}_\mathrm{KL}$) between the prior and each posterior, which will help us determine the number of rounds of LSBI required to obtain convergent results. The posterior distribution is also computed using nested sampling for comparison.

\begin{figure}[b]
\hspace{-4mm}
\includegraphics[width=0.5\textwidth]{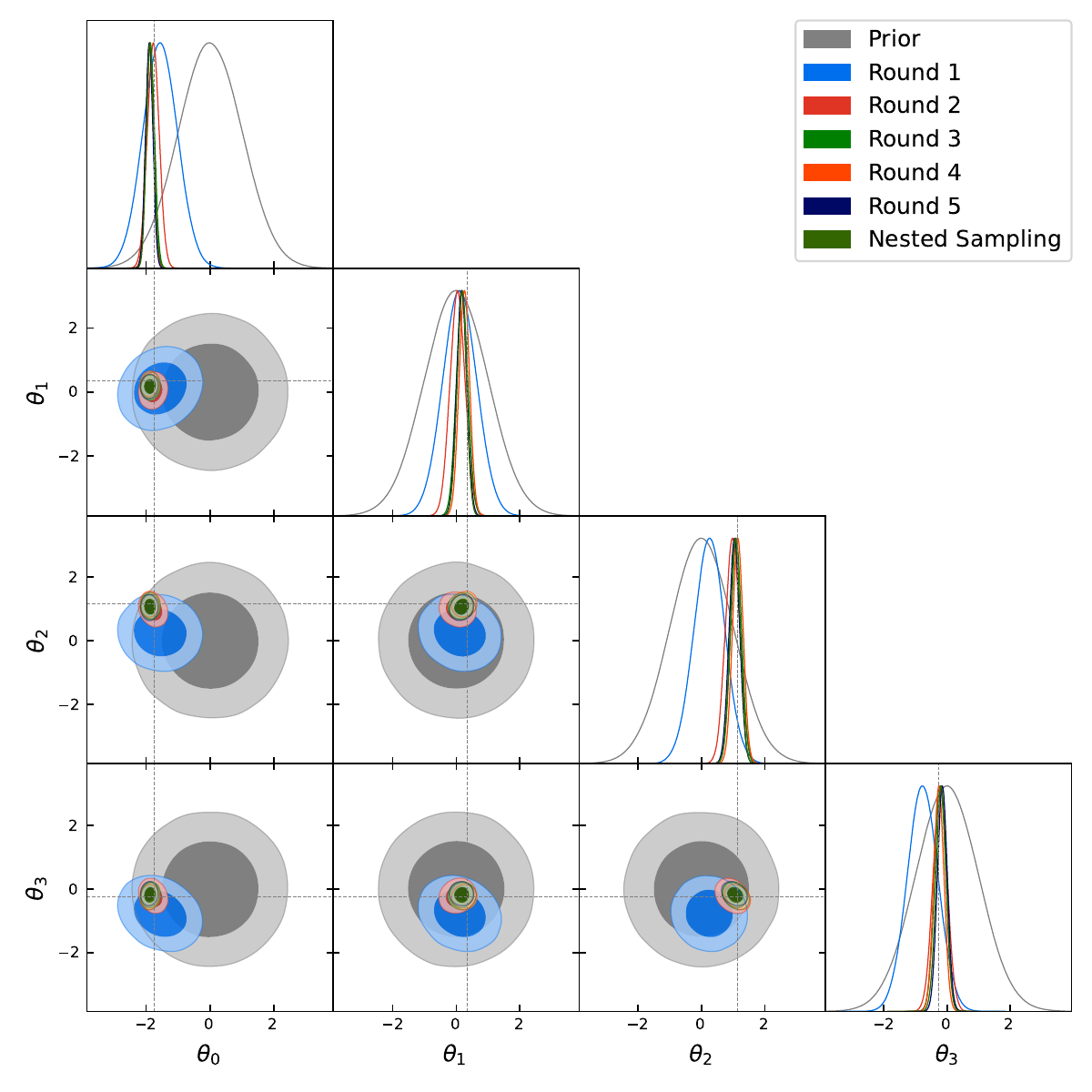}
\caption{\label{fig:quadratic}
Prior and posterior distributions on the parameters for a 50-dimensional dataset described by a non-linear 4-parameter model with Gaussian error. Each round corresponds to the output of LSBI after $k= 2500$ simulations, where the sampling distribution of the parameter sets $\{\theta^{(i)}\}$ is the posterior of the previous round. The result of nested sampling is also shown. The dashed lines indicate the values of the `real' parameters $\theta^\ast$.} 
\end{figure}

\begin{figure}[b]
\hspace{-4mm}
\includegraphics[width=0.45\textwidth]{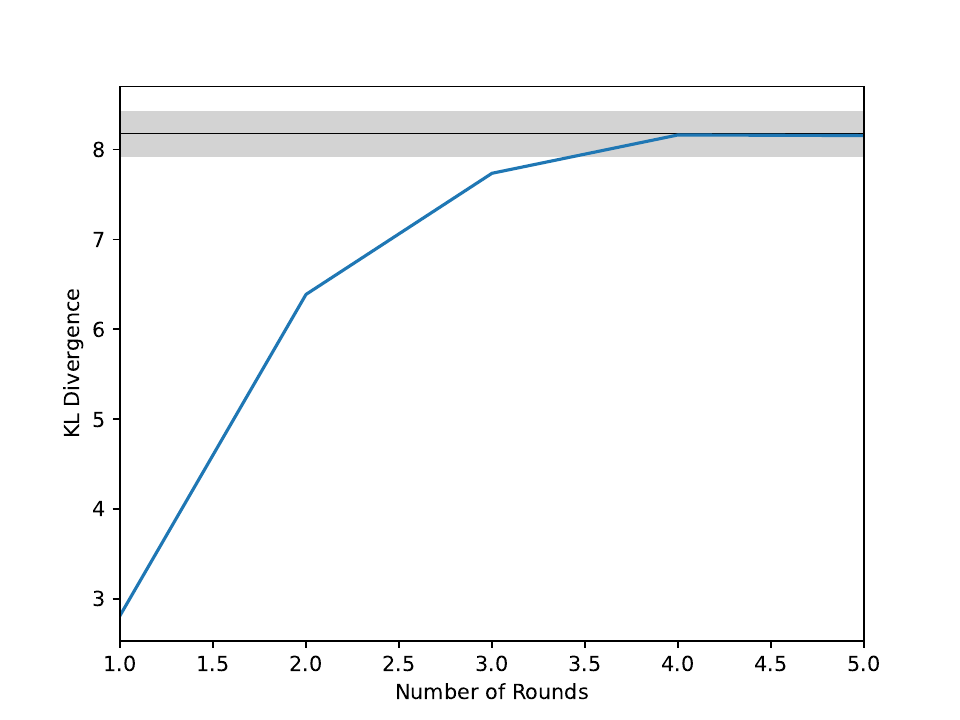}
\caption{\label{fig:quadratic_dkl}
$\mathcal{D}_\mathrm{KL}$ between the prior and posterior for each round of Sequential LSBI for the data displayed in Figure \ref{fig:quadratic} The black line corresponds to the value computed via nested sampling; the estimated error is also shown as a gray band.
}
\end{figure}

Figure~\ref{fig:quadratic} illustrates the outcomes of these simulations. The first iteration of LSBI indeed produces an excessively broad posterior, which subsequent iterations rapidly improve upon. Figure \ref{fig:quadratic_dkl} confirms that after four iterations, the Kullback-Leibler divergence between the prior and the LSBI posterior converges to that calculated via nested sampling, with no appreciable discrepancy as expected for Gaussian noise.

\begin{figure*}[t]
\includegraphics[width=0.3\textwidth]{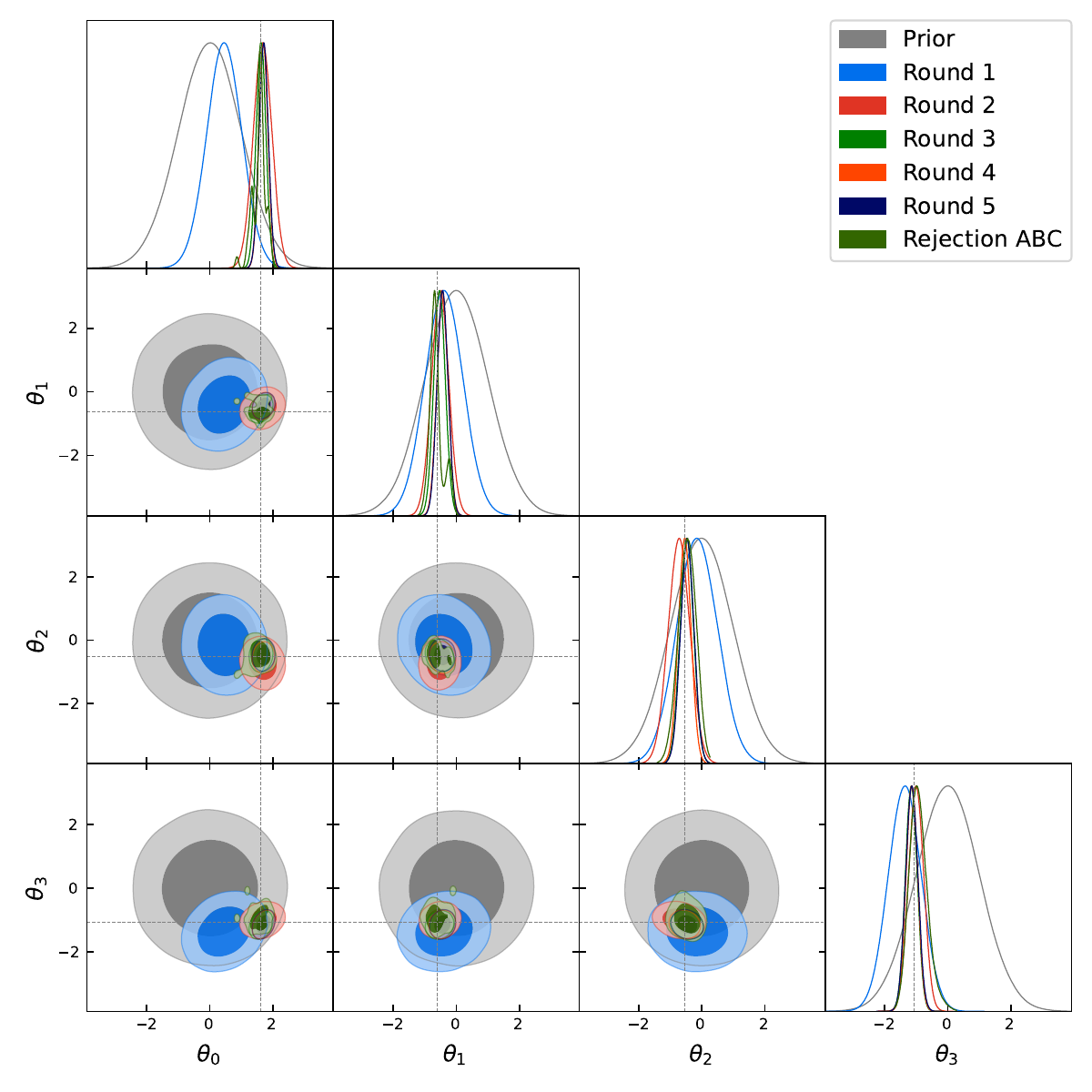}\qq
\includegraphics[width=0.3\textwidth]{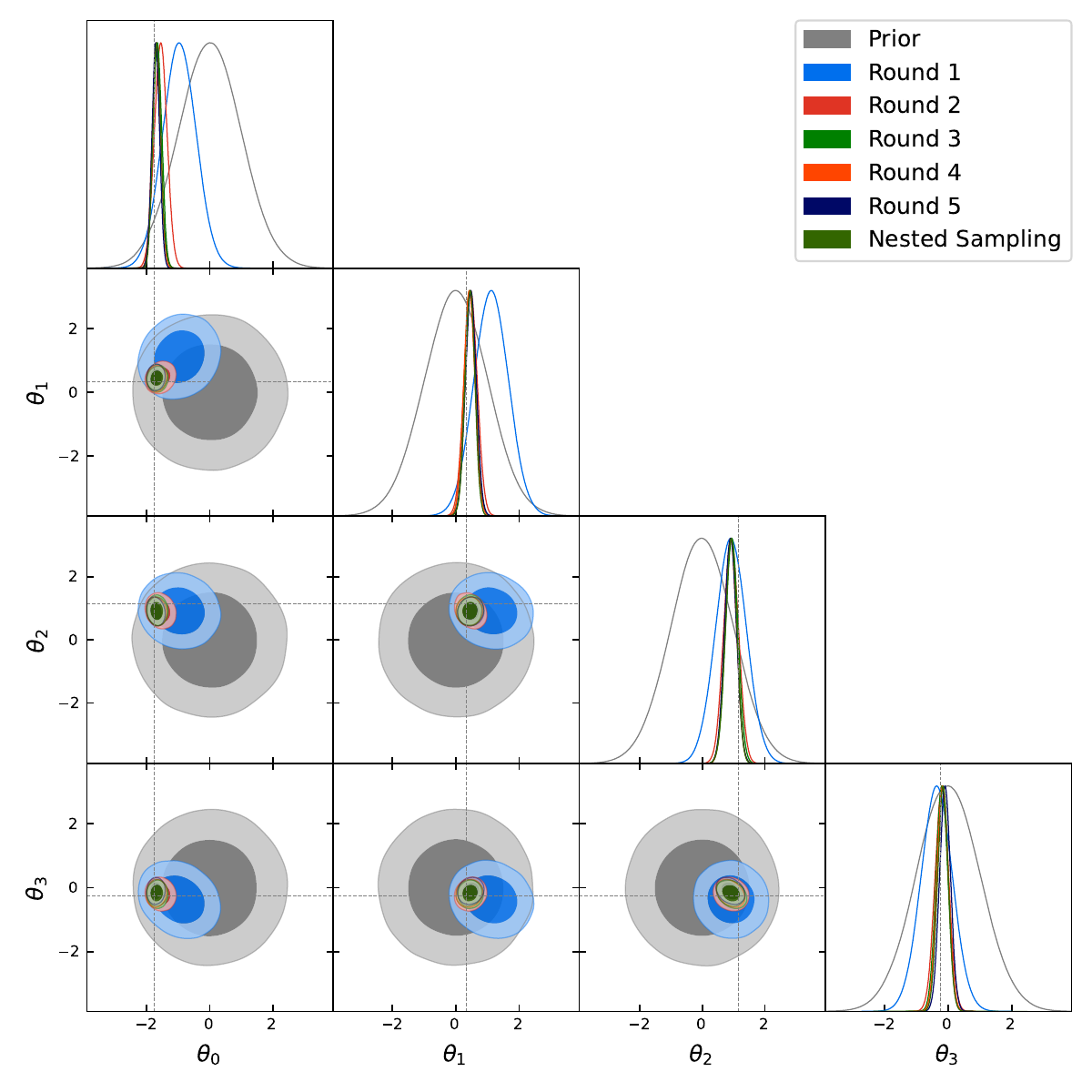}\qq
\includegraphics[width=0.3\textwidth]{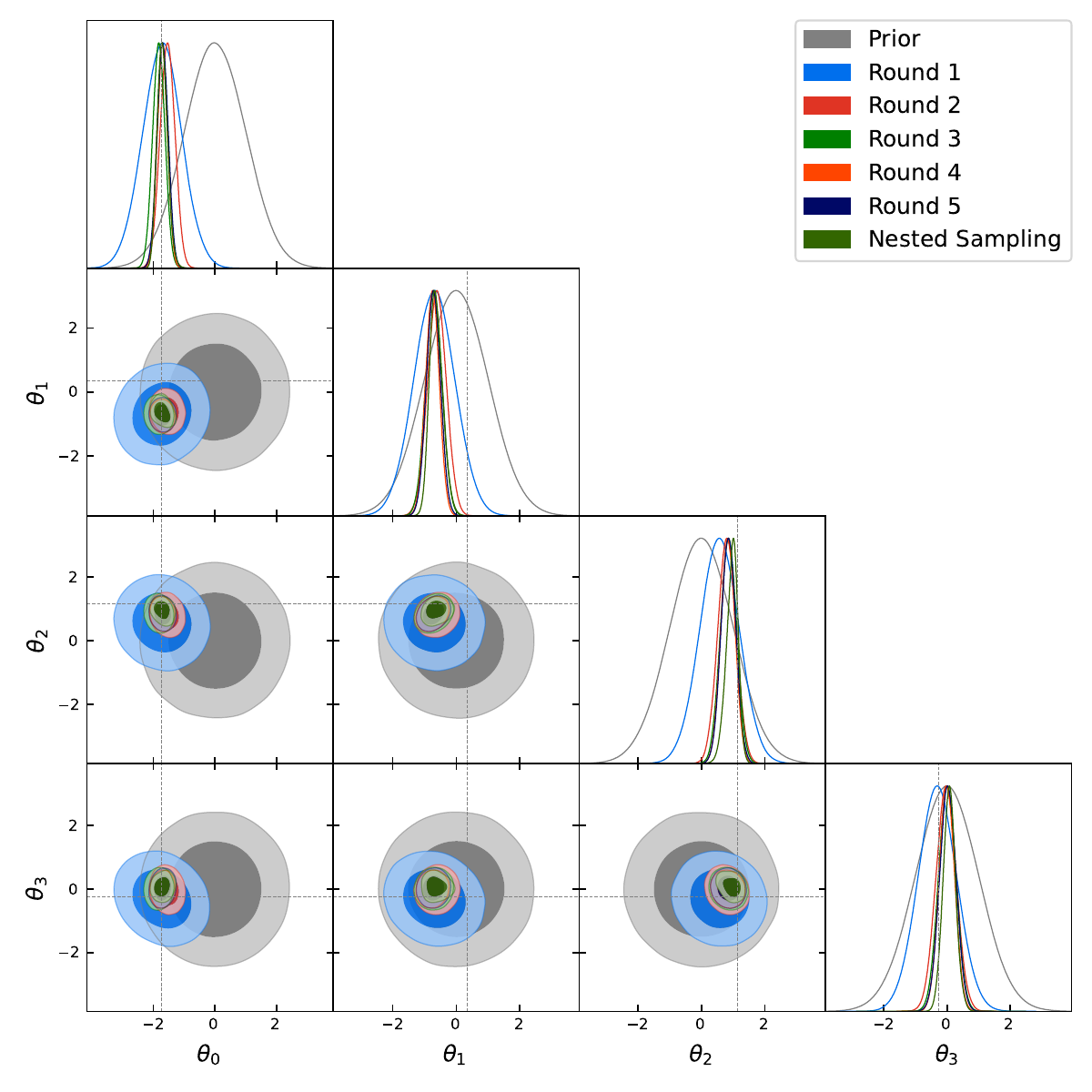}
\caption{\label{fig:non_gaussian} Prior and posterior distributions on the parameters for a 50-dimensional dataset described by a linear 4-parameter model and non-Gaussian error with $\sigma\approx 0.5$. The posteriors are computed from $k = 106,\ 500,\ 2500,$ and $10000$ samples drawn from the simulated likelihood. The dashed lines indicate the values of the underlying parameters $\theta^*$; (left) uniform noise; (centre) Student-$t$ noise; (right) asymmetric Laplacian noise. The posterior distribution is also computed using nested sampling for the Laplacian and Student-$t$ cases, whereas that of the uniform likelihood is obtained through an Approximate Bayesian Computation (rejection ABC).}
\end{figure*}

\begin{figure*}[t]
\includegraphics[width=0.3\textwidth]{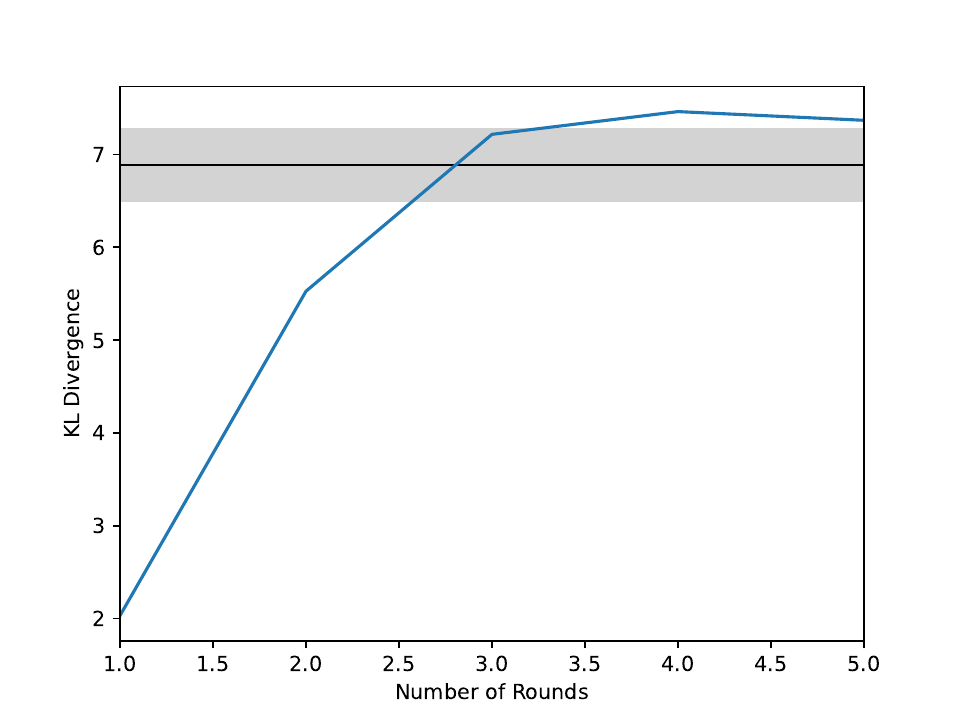}\qq
\includegraphics[width=0.3\textwidth]{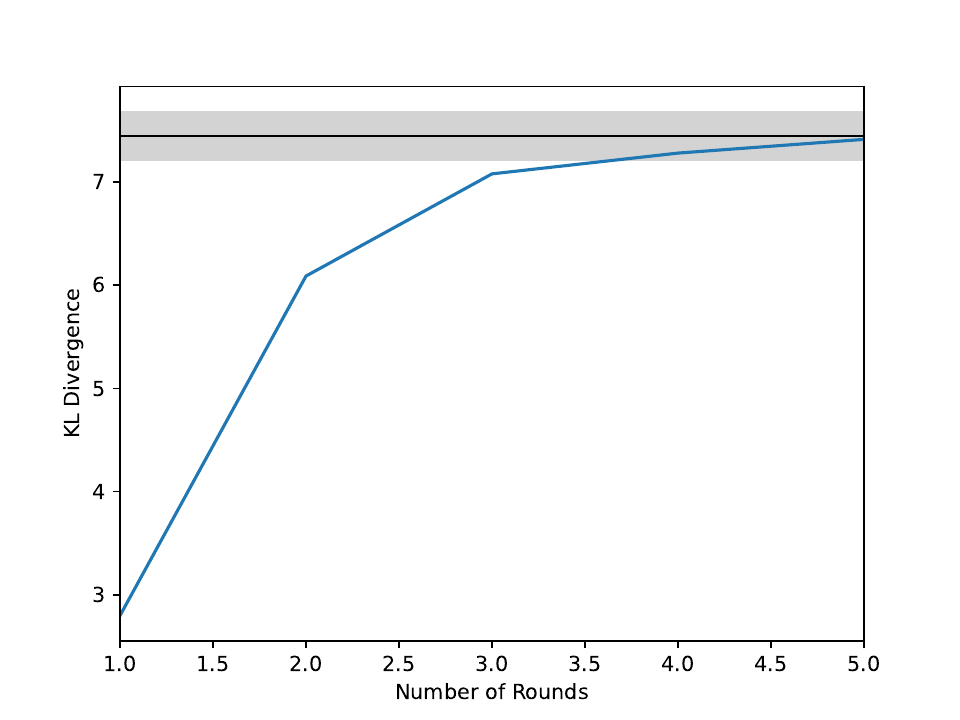}\qq
\includegraphics[width=0.3\textwidth]{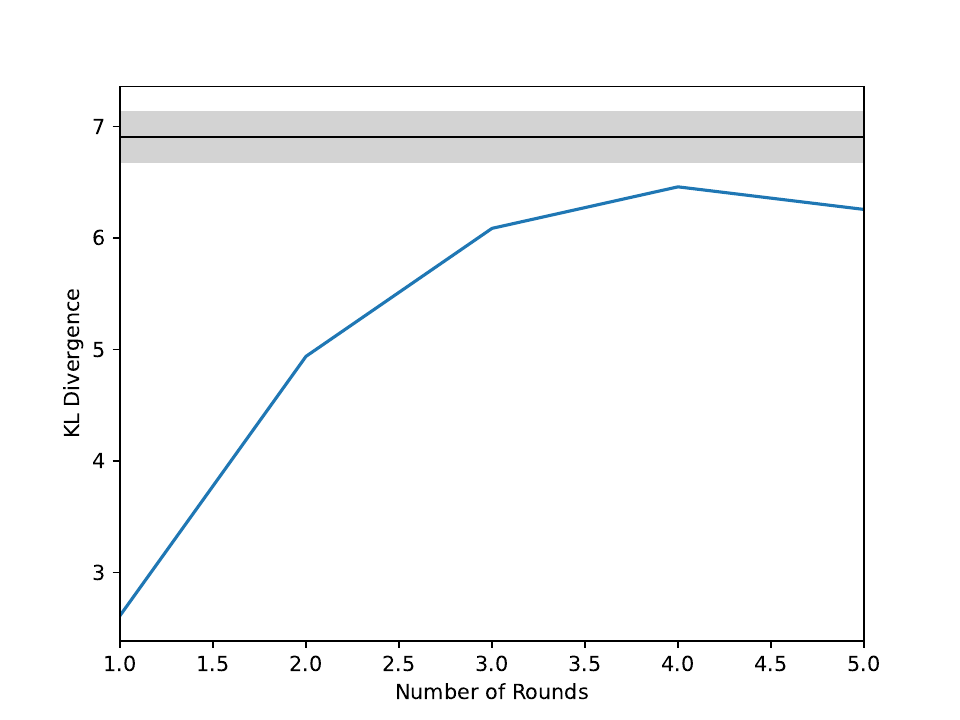}
\caption{\label{fig:non_gaussian_dkl} Kullback-Leibler divergence between the prior and posterior on the parameters as a function of the number of simulations; (left) uniform noise; (center) Student-$t$ noise; (right) asymmetric Laplacian noise. The black line corresponds to the value computed via nested sampling / rejection ABC; the estimated error is also shown as a gray band.}
\end{figure*}

In addition, we test the performance of LSBI for non-Gaussian noise shapes. The cases considered are uniform, Student-$t$ with $d+2$ degrees of freedom, and asymmetric Laplacian noise. The model is also given by Eq. \ref{eq:model}, and the uncertainty in these distributions is defined directly in terms of the model covariance $C$.  The posterior distribution is also computed using nested sampling for the Laplacian and Student-$t$ cases, whereas that of the uniform likelihood is obtained through an Approximate Bayesian Computation (rejection ABC).

The one- and two-dimensional LSBI posteriors for the models with non-Gaussian error are shown in Figures~\ref{fig:non_gaussian} and \ref{fig:non_gaussian_dkl}. The results demonstrate that the posteriors converge to a stable solution after approximately 4 rounds of sequential LSBI. Furthermore, the final $\mathcal{D}_\mathrm{KL}$ between the prior and posterior distributions approaches the values obtained using nested sampling / rejection ABC methods. Nevertheless, the distributions show some discrepancy, illustrating the fact that non-Gaussian noise can affect the accuracy of LSBI. The results are less satisfactory for Laplacian noise, as the $\mathcal{D}_\mathrm{KL}$ does not converge to a value within the error bars of the nested-sampling estimation. On the other hand, the lower value of the $\mathcal{D}_\mathrm{KL}$ estimated via rejection ABC for uniform noise compared to the LSBI values after round 3 is probably due to the inaccuracy of ABC as a posterior estimation method.

We note that the distributions considered here have well-defined first and second moments, so they can be approximated by a Gaussian. Although not considered in this paper, there exist distributions with an undefined covariance, such as the Cauchy distribution (Student-$t$ with one degree of freedom). In these cases, it has been checked that LSBI fails to predict a posterior, instead returning the original prior.

\subsection{The CMB Power Spectrum } \label{sec:results}

In this section, we test the performance of LSBI on a pseudo-realistic cosmological dataset. In the absence of generative Planck likelihoods, we produce the simulations through CosmoPowerJAX \cite{piras2023cosmopower,spurio2022cosmopower}, a library of machine learning-based inference tools and emulators. In particular, we use the $\texttt{cmb\_tt}$ probe, which takes as input the six $\mathrm{\Lambda}$CDM parameters: the Hubble constant $H_0$, the densities of baryonic matter $\Omega\txt{b}h^2$ and cold dark matter $\Omega\txt{c}h^2$, where $h = H_0/100$km\,s$^{-1}$\, Mpc$^{-1}$, the re-ionization optical depth $\tau$,  and the amplitude $A\txt{s}$ and slope $n\txt{s}$ of the initial power spectrum of density perturbations.
The output is the predicted CMB temperature power spectrum
\begin{equation}
	\mathcal{C}_\ell = \f{1}{2\ell+1}\sum_{m=-\ell}^\ell |a_{\ell,m}|^2, \qq 2\leqslant\ell\leqslant2058
\end{equation}
where $a_{\ell,m}$ are the coefficients of the harmonic expansion of the CMB signal. To this data, we add the standard scaled $\chi^2$ noise
\begin{equation}\label{eq:chi2noise}
	\f{2\ell+1}{C_\ell+N_\ell}\mathcal{C_\ell}\ \sim\  \chi^2(2\ell+1).
\end{equation}

We apply several rounds of Sequential LSBI, each drawing $k=10^4$ simulations from the emulator, but keep $N=100$ since a larger number is computationally unmanageable without parallelization. The parameter samples are drawn from the prior displayed in Eqs.~\ref{eq:prior_CMB_1} and \ref{eq:prior_CMB_2} and, as before, the observed data is generated by running the simulator on a known set of parameters $\theta^\ast$. The posterior is also obtained by nested sampling with \texttt{dynesty}.

\begin{figure*}[p!]
\includegraphics[width=0.9\textwidth]{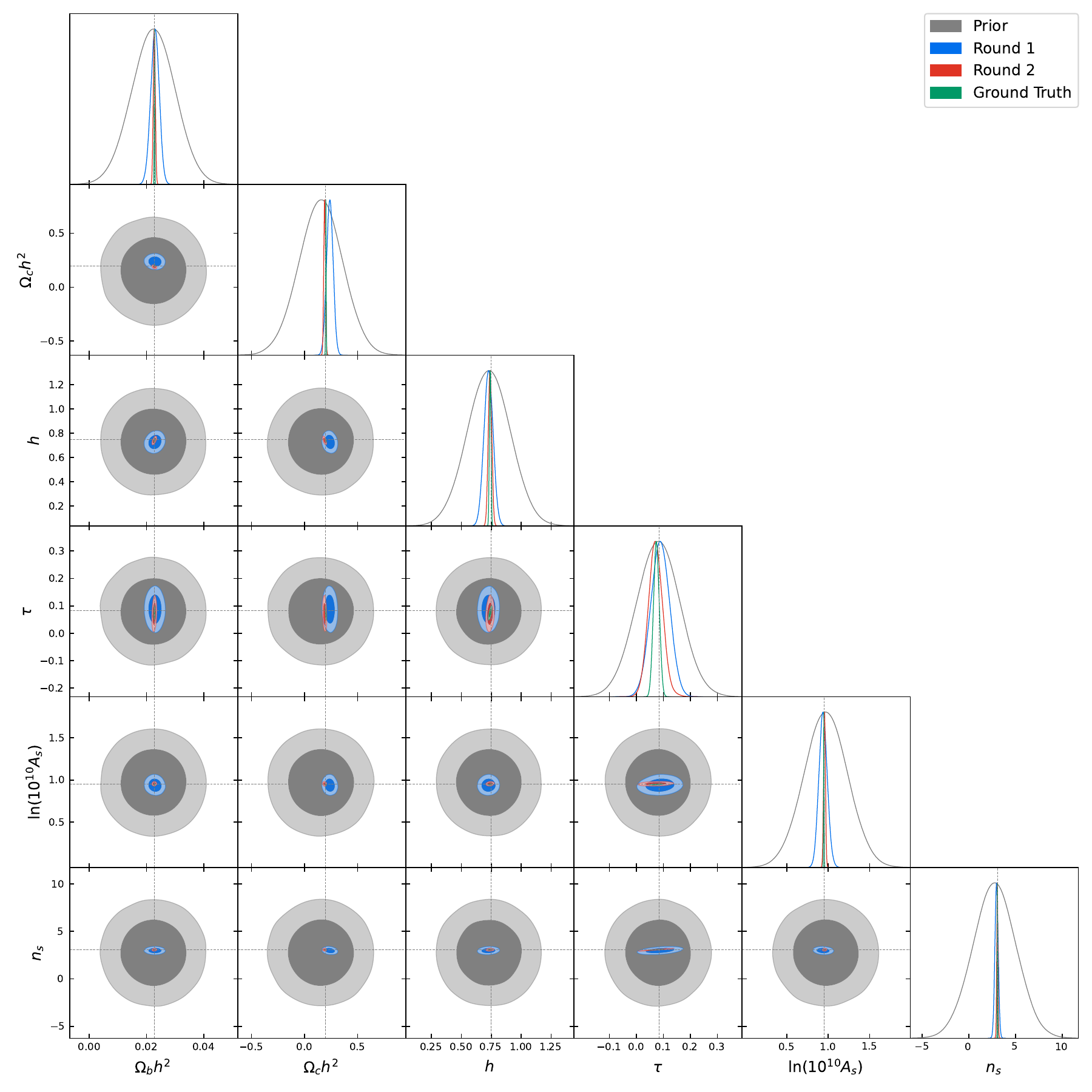}
\caption{\label{fig:CMB_posterior_1}
The plot displays the two-dimensional posterior distributions given by the first two rounds of sequential LSBI, where each round corresponds to the output of LSBI after $k= 10^4$ simulations. The prior distribution and the result of nested sampling on the dataset (labeled "Ground Truth") are also shown. The dashed lines indicate the values of the `real' parameters $\theta^\ast$.}
\end{figure*}

\begin{figure*}
\includegraphics[width=0.9\textwidth]{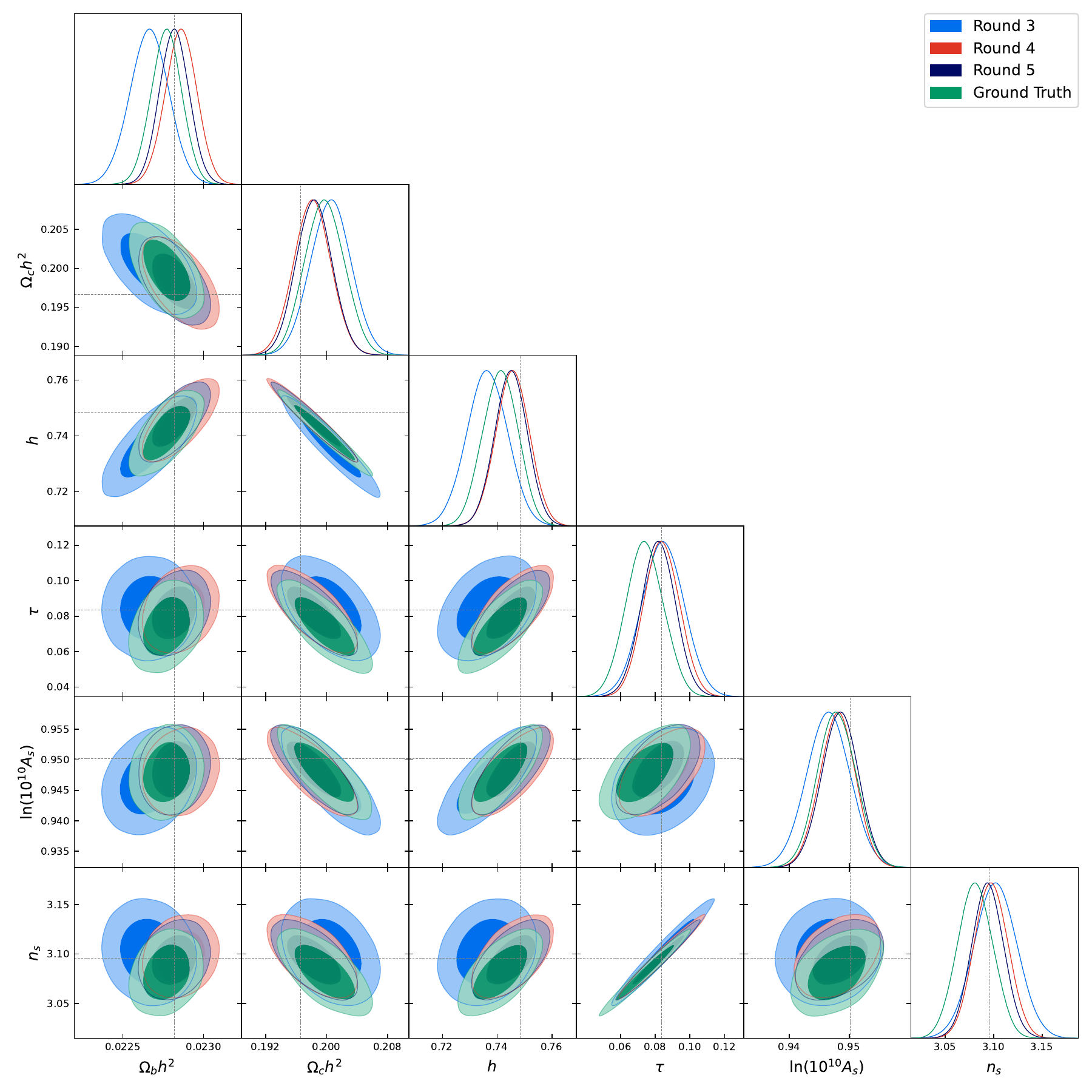}
\caption{\label{fig:CMB_posterior_2}
The plot displays the two-dimensional posterior distributions given by rounds three through five of sequential LSBI, together with the result of nested sampling on the dataset (labeled "Ground Truth"). The dashed lines indicate the values of the `real' parameters $\theta^\ast$.
}
\end{figure*}

The output of this calculation is shown in Figs. \ref{fig:CMB_posterior_1} and \ref{fig:CMB_posterior_2}. The first figure displays the prior and rounds 1 and 2, while the second shows rounds 3 to 5; in both cases the nested sampling result is shown.  It can be noted by eye that the posterior coincides well with the result of nested sampling after four to five rounds of LSBI. This suggests that, although the CMB power spectrum is not well approximated by a linear model at first instance, sequential LSBI succeeds at yielding a narrow sampling distribution about $\theta^\ast$,  thus iteratively approximating the correct posterior.  

\begin{figure}[t]
\hspace{-4mm}
\includegraphics[width=0.45\textwidth]{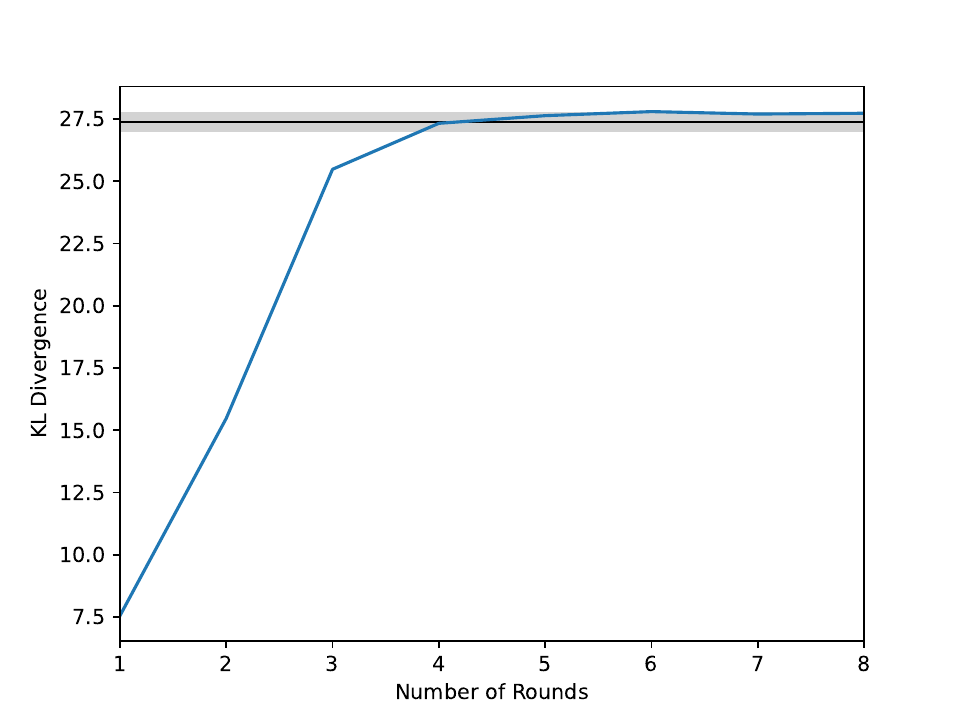}
\caption{\label{fig:CMB_posterior_dkl}
$\mathcal{D}_\mathrm{KL}$ between the prior and posterior for each round of Sequential LSBI for the simulated CMB data, displayed in Figures \ref{fig:CMB_posterior_1} and \ref{fig:CMB_posterior_2}. The black line corresponds to the value computed via nested sampling. The estimated error nested sampling is also shown as a gray band.
}
\end{figure}

Figure \ref{fig:CMB_posterior_dkl} displays the evolution of the KL divergence up to 8 rounds of LSBI. This result provides further evidence that sequential LSBI converges after $\mathcal{O}(1)$ rounds, thus keeping the total number of simulations within $\mathcal{O}(10^4)$. Nevertheless, we note that the $ \mathcal{D}_\mathrm{KL}$ is slightly overestimated, suggesting that the LSBI posterior overestimates the true distribution to a small extent. Reproducing this calculation with smaller values of $N$, we have noticed that the overconfidence increases as $N$ is decreased. Therefore, as discussed in Sections \ref{sec:LSBI} and  \ref{sec:seqLSBI}, the choice of $N$ must be large enough to guarantee the convergence of the integral in Eq. \ref{eq:posterior_est}. In general we recommend a value at least of order $10^3$, but as evidenced by Figure \ref{fig:CMB_posterior_dkl}, a smaller value may still yield accurate results.

\section{Conclusion}

In this paper, we have developed the theoretical framework of Linear Simulation-based Inference (LSBI), an application of likelihood-free inference where the model is approximated by a linear function of its parameters and the noise is assumed to be Gaussian with zero mean. Under these circumstances, it is possible to obtain analytical expressions for the posterior distributions of hyper-parameters of the linear likelihood, given a set of samples $S = (\theta^{(i)},D^{(i)})$, where $D^{(i)}$ are obtained by running simulations on the $\theta^{(i)}$. These parameter samples can be drawn from a distribution other than the prior, so we can exploit this to sequentially approximate the posterior in the vicinity of the observed data. 

The analysis of the toy models in Section \ref{sec:toy_models} illustrates the extent of the resilience of LSBI to deviations from its assumptions. When the error is non-Gaussian, LSBI can still yield accurate estimates, although not universally so, and sequential LSBI provides a way to effectively treat non-linear models.  Furthermore, its application to the pseudo-realistic model for the CMB power spectrum demonstrates that it is possible to obtain significant information gain and generate posteriors that agree with the underlying parameters while maintaining explainability and intellectual oversight.   We also find that convergence is achieved after $\mathcal{O}(10^4)$ simulations, competitive with state-of-the-art neural density estimation methods \cite{papamakarios2016fast,alsing2019fast,cole2022fast}. 

Further efforts should be directed towards testing LSBI on more realistic examples, such as the CMB with foregrounds, Baryon Acoustic Oscillations (BAO), or supernovae surveys.
In addition, extending this analysis to Gaussian-mixture models may be helpful in better approximating non-Gaussian and multimodal likelihoods.

\section{Materials}

The source code for the data and plots shown in this paper can be found at \href{https://github.com/ngm29/astro-lsbi/tree/main}{https://github.com/ngm29/astro-lsbi/tree/main}.

\vfill

\clearpage

\appendix

\begin{widetext}

\section{LSBI with Conjugate Priors} \label{ap:A}

Consider a simulator $\S_\M$ which emulates a model $\M$ that can be approximated linearly. We ignore the values of the hyper-parameters $m$, $M$, and $C$, but we may infer them by performing $k$ \emph{independent} simulations $S = \{(D_i, \theta_i)\}$, where the $\{\theta_i\}$ may be drawn from an arbitrary distribution. Defining the sample covariances and means for the data and parameters,
\begin{gather}
    \Theta 	= \tfrac{1}{k}\sum_i (\theta_i-\m{\theta})(\theta_i-\m{\theta})^\T, \qquad
    \Delta 	= \tfrac{1}{k}\sum_i ( D_i-\m{ D})( D_i-\m{D})^\T, \\
    \Psi 	= \tfrac{1}{k}\sum_i ( D_i-\m{D})(\theta_i-\m{\theta})^\T, \qquad
    \m{D} 	= \tfrac{1}{k}\sum_i  D_i, \qquad
    \m{\theta} = \tfrac{1}{k}\sum_i \theta_i, 
\end{gather}
then, after some algebra, the joint likelihood for the simulations can be written as:
\begin{align*}
	\log p(\{D_i\}|\{\theta_i\},m,M,C) 
		&\equiv \textstyle\sum_i \log p(D_i|\theta_i,m,M,C)								\\
		& = -\tfrac{k}{2}\log|2\pi C|
			-\tfrac{1}{2}\textstyle\sum_i ( D_i - m - M\theta_i)^\T C^{-1}( D_i - m - M\theta_i) \\
		& = -\tfrac{k}{2}\log|2\pi C|
			-\half\lrb{ m- (\m D-M\m\theta) }^\T (C/k)^{-1}\lrb{ m- (\m D-M\m\theta)} 	\\
		&\qquad\qquad\qquad\q\! \textstyle 
			-\frac{k}{2} \tr\lrp{ \Theta (M-\Psi\Theta^{-1})^\T C^{-1}(M-\Psi\Theta^{-1})}\\
		&\qquad\qquad\qquad\q\! \textstyle  
			- \frac{k}{2} \tr\lrp{(\Delta -  \Psi \Theta^{-1}\Psi^\T)C^{-1}}  .
\end{align*}

\noindent Using the previous result, we can infer via Bayes' theorem
$$
	p(m|M,C,S) = \frac{p\lrp{\{D_i\}|\{\theta_i\},m,M,C}}{p\lrp{\{D_i\}|\{\theta_i\},M,C}}\cdot p\lrp{m|M,C,\{\theta^{(i)}\}}.
$$
We choose the conjugate prior $m|M,C,\{\theta^{(i)}\}  \sim  \N(0,C)$, giving the posterior
\begin{equation} \label{model_m_pdf}
	p(m|M,C,S) = \frac{1}{\sqrt{(2\pi)^d|\frac{C}{k+1}|}}
		\exp\lrp{-\frac{1}{2}\lrb{ m- \tfrac{k}{k+1}(\m D-M\m\theta) }^\T \lrb{C/(k+1)}^{-1}\lrb{ m- \tfrac{k}{k+1}(\m D-M\m\theta)}}.
\end{equation}
Hence, we find a new distribution, with $m$ integrated out (the running evidence),
\begin{align*} 
	\log p(\{D_i\}|\{\theta_i\},M,C)
	&=-\tfrac{k}{2}\log|2\pi C|
			-\tfrac{d}{2}\log(k+1)\\
		&\qq \textstyle 
			-\tfrac{1}{2}\tfrac{k}{k+1}(\m D -M\m\theta)^\T C^{-1}(\m D -M\m\theta)\\
		&\qq \textstyle 
			-\frac{k}{2} \tr\lrp{ \Theta (M-\Psi\Theta^{-1})^\T C^{-1}(M-\Psi\Theta^{-1})}\\
		&\qq \textstyle  
			- \frac{k}{2} \tr\lrp{(\Delta -  \Psi \Theta^{-1}\Psi^\T)C^{-1}}\\
	&=-\tfrac{k}{2}\log|2\pi C|
			-\tfrac{d}{2}\log(k+1)\\
		&\qq \textstyle 
			-\frac{k}{2} \tr\lrp{ \Theta_* (M-\n\Psi\Theta_*^{-1})^\T C^{-1}(M-\n\Psi\Theta_*^{-1})}\\
		&\qq \textstyle  
			- \frac{k}{2} \tr\lrp{(\n\Delta -  \n\Psi \Theta_*^{-1}\n\Psi^\T)C^{-1}},
\end{align*}
where 	
\begin{align}
	\n\Theta \equiv \Theta + \tfrac{1}{k+1}\m\theta\,\m\theta^\T,\qq
	\n\Psi \equiv \Psi + \tfrac{1}{k+1}\m D\,\m\theta^\T,\qq
	\n\Delta\equiv \Delta + \tfrac{1}{k+1}\m D\,\m D^\T.
\end{align}
 A second application of Bayes' theorem gives
$$
	p(M|C,S) = \frac{p\lrp{\{D_i\}|\{\theta_i\},M,C}}{p\lrp{\{D_i\}|\{\theta_i\},C}}\c p\lrp{M|C,\{\theta^{(i)}\}}.
$$
We choose the conjugate prior $M|C,\{\theta^{(i)}\}  \sim  \N(0,C,\Theta^{-1})$, giving the posterior
\begin{equation} 
	p(M|C,S) = \frac{1}{\sqrt{(2\pi)^{nd}|\frac{C}{k}|^n|\Theta^{-1}_*|^{d}}}
		\exp\lrp{-\frac{1}{2}\tr\lrB{\Theta_* \lrb{M-\n\Psi\Theta_*^{-1}}^\T \lrp{\frac{C}{k}}^{-1}\lrb{M-\n\Psi\Theta_*^{-1}}}},
\end{equation}
where 
\begin{equation}
	\Theta_* \equiv \n\Theta + \tfrac{1}{k}\Theta = \tfrac{k+1}{k}\Theta + \tfrac{1}{k+1}\m\theta\,\m\theta^\T.
\end{equation}
The running evidence after marginalization over $M$ is then,
\begin{align*} 
	\log p(\{D_i\}|\{\theta_i\},C)
	&=-\tfrac{k}{2}\log|2\pi C| -\tfrac{d}{2}\log(k+1) -\tfrac{dn}{2}\log k+\tfrac{d}{2}\log|\Theta_*^{-1}\Theta|\\
		&\qq \textstyle 
			- \frac{k}{2} \tr\lrB{\lrp{\n\Delta -  \n\Psi \lrb{\Theta_*^{-1}+\n\Theta^{-1}- \Theta^{-1}}\n\Psi^\T}C^{-1}}.
\end{align*}

\noindent A third and final application of Bayes' theorem
$$
	p(C|S) = \frac{p(\{D_i\}|\{\theta_i\},C)}{p(\{D_i\}|\{\theta_i\})}\c p\lrp{C,\{\theta^{(i)}\}}
$$
with conjugate prior $C|\{\theta^{(i)}\}  \sim  \W^{-1}(C_0,\nu_0)$, gives the posterior
\begin{equation}
	p(C|S) =\frac{|C|^{-(\nu+d +1)/2}}{N(S)}
			\exp\lrp{-\frac{1}{2}\tr\lrB{\lrb{k\lrp{\n\Delta -  \n\Psi \lrb{\Theta_*^{-1}+\n\Theta^{-1}- \Theta^{-1}}\n\Psi^\T}+C_0}C^{-1}}}
\end{equation}
with $\nu = \nu_0 + k$ and $\nu_0>d$, and
\begin{equation*}
	N(S)= 2^{\nu d/2}  \Gamma_d[\tfrac{\nu}{2}]\times\left|k\lrp{\n\Delta -  \n\Psi \lrb{\Theta_*^{-1}+\n\Theta^{-1}- \Theta^{-1}}\n\Psi^\T}+C_0\right|^{-\nu/2}
\end{equation*}
. The running evidence is then
\begin{align*}
	\log p(\{D_i\}|\{\theta_i\}) &\equiv  \frac{p(\{D_i\}|\{\theta_i\},C)}{p(C|S)}\cdot p(C) \\
	&=-\tfrac{kd}{2}\log \pi -\tfrac{d}{2}\log(k+1) -\tfrac{dn}{2}\log k+\log(\Gamma_d[\tfrac{\nu}{2}]/\Gamma_d[\tfrac{\nu_0}{2}])\\
		&\qq \textstyle 
			+\tfrac{d}{2}\log|\n\Theta^{-1}\Theta|-\tfrac{1}{2}\log\lrB{\left|k\lrp{\Delta -  \Psi \lrb{\Theta_*^{-1}+\n\Theta^{-1}- \Theta^{-1}}\Psi^\T}+C_0\right|^\nu\big/|C_0|^{\nu_0}}.
\end{align*}

Finally, we can compute the total evidence for the simulations, where we assume that the parameter samples have been drawn from a Gaussian, $\theta^{(i)}\sim\N(\m \theta, \Theta)$
\begin{align*}
	\log p(S) &= \log p(\{D_i\}|\{\theta_i\}) + \log p(\{\theta_i\})\\
	&=  \log p(\{D_i\}|\{\theta_i\}) - \tfrac{k}{2}\log|2\pi\Theta|-\tfrac{1}{2}nk\\[2mm]
	&=-\tfrac{kd}{2}\log \pi -\tfrac{d}{2}\log(k+1) -\tfrac{dn}{2}\log k+\log(\Gamma_d[\tfrac{\nu}{2}]/\Gamma_d[\tfrac{\nu_0}{2}])-\tfrac{1}{2}nk\\
		&\qq \textstyle 
			+\tfrac{d}{2}\log|\n\Theta^{-1}\Theta|- \tfrac{k}{2}\log|2\pi\Theta|-\tfrac{1}{2}\log\lrB{\left|k\lrp{\Delta -  \Psi \lrb{\Theta_*^{-1}+\n\Theta^{-1}- \Theta^{-1}}\Psi^\T}+C_0\right|^\nu\big/|C_0|^{\nu_0}}.
\end{align*}

\section{Model Comparison and Data Compression} \label{ap:C}
If the implicit likelihood is inferred  for a different model $\M'$, the Bayes' ratio between the two models may be calculated,
\begin{equation}
	\mathcal{B} = \f{\langle p(D\txt{obs}|\M)\rangle_{m,M,C}}{\langle p(D\txt{obs}|\M')\rangle_{m',M',C'}}
\end{equation}

However, this calculation requires a $d\times d$ matrix to be inverted (see Eq. \ref{eq:evidence}), which scales as $\mathcal{O}(d^\alpha \times N)$, where $2<\alpha\leqslant3$ depending on the algorithm used. To increase the computational efficiency, Alsing et.\ al.\ \cite{alsing2018generalized, alsing2018massive} and Heavens et.\ al.\ \cite{heavens2000massive} remark that for a homoscedastic Gaussian likelihood, the data may be mapped into a set of $n$ summary statistics via the linear compression
\begin{equation} \label{eq:compression}
	D \ \mapsto \ M^\T C^{-1}(D-m)
\end{equation}
without information loss. In our case, the Gaussian likelihood is an approximation, so we may only claim lossless compression to the extent that the approximation is good. The advantage of this method is that $C^{-1}$ may be drawn directly from the distribution
\begin{align}
	C^{-1}|S &\sim \W\lrp{[(k-1)(\Delta -  \Psi \Theta^{-1}\Psi^\T)]^{-1}, \nu }
\end{align}
and thus, save for one-time inversion of the scale matrix, the complexity of the compression is no more than $\mathcal{O}(n\times d^2\times N)$.

We propose a slightly different data compression scheme,
\begin{equation}
	 x = \Gamma M^\T C^{-1}(D-m), \q \Gamma \equiv (M^\T C^{-1}M)^{-1},
\end{equation} where $\Gamma$ can be computed in $\mathcal{O}(n^2\times d^2 \times N)$ time. We substitute into Eq. \ref{eq:evidence} to find
\begin{align*}
	x|\M &\sim \mathcal{N}(\mu,\Sigma + \Gamma).
\end{align*}
The matrix $\Sigma+\Gamma$ is only $n\times n$, so if we assume that $n\ll d$, this represents a potential reduction in computational complexity.

\section{Priors} \label{ap:B}

\noindent Prior for the $\Lambda$CDM parameters, $\theta\txt{CMB} = (\Omega\txt{b}h^2, \Omega\txt{c}h^2, \tau, \ln(10^{10}A\txt{s}), n\txt{s},H_0)$
\begin{align}
\mu\txt{CMB} &= (2.22\t10^{-2}, 0.120, 6.66\t10^{-2}, 3.05, 0.964\t10^{-1}, 67.3),
\label{eq:prior_CMB_1}\\[2mm]
\Sigma\txt{CMB} &= \text{diag}(1.05\t10^{-3}, 8.28\t10^{-3}, 3.47\t10^{-2}, 1.47\t10^{-1}, 2.64\t10^{-2}, 3.38). \label{eq:prior_CMB_2}
\end{align}
\vfill
\pagebreak
\end{widetext}

\bibliography{main.bib}

\providecommand{\noopsort}[1]{}\providecommand{\singleletter}[1]{#1}%
\begin{thebibliography}{30}%
\makeatletter
\providecommand \@ifxundefined [1]{%
 \@ifx{#1\undefined}
}%
\providecommand \@ifnum [1]{%
 \ifnum #1\expandafter \@firstoftwo
 \else \expandafter \@secondoftwo
 \fi
}%
\providecommand \@ifx [1]{%
 \ifx #1\expandafter \@firstoftwo
 \else \expandafter \@secondoftwo
 \fi
}%
\providecommand \natexlab [1]{#1}%
\providecommand \enquote  [1]{``#1''}%
\providecommand \bibnamefont  [1]{#1}%
\providecommand \bibfnamefont [1]{#1}%
\providecommand \citenamefont [1]{#1}%
\providecommand \href@noop [0]{\@secondoftwo}%
\providecommand \href [0]{\begingroup \@sanitize@url \@href}%
\providecommand \@href[1]{\@@startlink{#1}\@@href}%
\providecommand \@@href[1]{\endgroup#1\@@endlink}%
\providecommand \@sanitize@url [0]{\catcode `\\12\catcode `\$12\catcode `\&12\catcode `\#12\catcode `\^12\catcode `\_12\catcode `\%12\relax}%
\providecommand \@@startlink[1]{}%
\providecommand \@@endlink[0]{}%
\providecommand \url  [0]{\begingroup\@sanitize@url \@url }%
\providecommand \@url [1]{\endgroup\@href {#1}{\urlprefix }}%
\providecommand \urlprefix  [0]{URL }%
\providecommand \Eprint [0]{\href }%
\providecommand \doibase [0]{https://doi.org/}%
\providecommand \selectlanguage [0]{\@gobble}%
\providecommand \bibinfo  [0]{\@secondoftwo}%
\providecommand \bibfield  [0]{\@secondoftwo}%
\providecommand \translation [1]{[#1]}%
\providecommand \BibitemOpen [0]{}%
\providecommand \bibitemStop [0]{}%
\providecommand \bibitemNoStop [0]{.\EOS\space}%
\providecommand \EOS [0]{\spacefactor3000\relax}%
\providecommand \BibitemShut  [1]{\csname bibitem#1\endcsname}%
\let\auto@bib@innerbib\@empty
\bibitem [{\citenamefont {Cranmer}\ \emph {et~al.}(2020)\citenamefont {Cranmer}, \citenamefont {Brehmer},\ and\ \citenamefont {Louppe}}]{cranmer2020frontier}%
  \BibitemOpen
  \bibfield  {author} {\bibinfo {author} {\bibfnamefont {K.}~\bibnamefont {Cranmer}}, \bibinfo {author} {\bibfnamefont {J.}~\bibnamefont {Brehmer}},\ and\ \bibinfo {author} {\bibfnamefont {G.}~\bibnamefont {Louppe}},\ }\bibfield  {title} {\bibinfo {title} {The frontier of simulation-based inference},\ }\href@noop {} {\bibfield  {journal} {\bibinfo  {journal} {Proceedings of the National Academy of Sciences}\ }\textbf {\bibinfo {volume} {117}},\ \bibinfo {pages} {30055} (\bibinfo {year} {2020})}\BibitemShut {NoStop}%
\bibitem [{\citenamefont {Rubin}(1984)}]{rubin1984bayesianly}%
  \BibitemOpen
  \bibfield  {author} {\bibinfo {author} {\bibfnamefont {D.~B.}\ \bibnamefont {Rubin}},\ }\bibfield  {title} {\bibinfo {title} {Bayesianly justifiable and relevant frequency calculations for the applied statistician},\ }\href@noop {} {\bibfield  {journal} {\bibinfo  {journal} {The Annals of Statistics}\ ,\ \bibinfo {pages} {1151}} (\bibinfo {year} {1984})}\BibitemShut {NoStop}%
\bibitem [{\citenamefont {Marjoram}\ \emph {et~al.}(2003)\citenamefont {Marjoram}, \citenamefont {Molitor}, \citenamefont {Plagnol},\ and\ \citenamefont {Tavar{\'e}}}]{marjoram2003markov}%
  \BibitemOpen
  \bibfield  {author} {\bibinfo {author} {\bibfnamefont {P.}~\bibnamefont {Marjoram}}, \bibinfo {author} {\bibfnamefont {J.}~\bibnamefont {Molitor}}, \bibinfo {author} {\bibfnamefont {V.}~\bibnamefont {Plagnol}},\ and\ \bibinfo {author} {\bibfnamefont {S.}~\bibnamefont {Tavar{\'e}}},\ }\bibfield  {title} {\bibinfo {title} {Markov chain monte carlo without likelihoods},\ }\href@noop {} {\bibfield  {journal} {\bibinfo  {journal} {Proceedings of the National Academy of Sciences}\ }\textbf {\bibinfo {volume} {100}},\ \bibinfo {pages} {15324} (\bibinfo {year} {2003})}\BibitemShut {NoStop}%
\bibitem [{\citenamefont {Sisson}\ \emph {et~al.}(2007)\citenamefont {Sisson}, \citenamefont {Fan},\ and\ \citenamefont {Tanaka}}]{sisson2007sequential}%
  \BibitemOpen
  \bibfield  {author} {\bibinfo {author} {\bibfnamefont {S.~A.}\ \bibnamefont {Sisson}}, \bibinfo {author} {\bibfnamefont {Y.}~\bibnamefont {Fan}},\ and\ \bibinfo {author} {\bibfnamefont {M.~M.}\ \bibnamefont {Tanaka}},\ }\bibfield  {title} {\bibinfo {title} {Sequential monte carlo without likelihoods},\ }\href@noop {} {\bibfield  {journal} {\bibinfo  {journal} {Proceedings of the National Academy of Sciences}\ }\textbf {\bibinfo {volume} {104}},\ \bibinfo {pages} {1760} (\bibinfo {year} {2007})}\BibitemShut {NoStop}%
\bibitem [{\citenamefont {Papamakarios}\ and\ \citenamefont {Murray}(2016)}]{papamakarios2016fast}%
  \BibitemOpen
  \bibfield  {author} {\bibinfo {author} {\bibfnamefont {G.}~\bibnamefont {Papamakarios}}\ and\ \bibinfo {author} {\bibfnamefont {I.}~\bibnamefont {Murray}},\ }\bibfield  {title} {\bibinfo {title} {Fast $\varepsilon$-free inference of simulation models with bayesian conditional density estimation},\ }\href@noop {} {\bibfield  {journal} {\bibinfo  {journal} {Advances in neural information processing systems}\ }\textbf {\bibinfo {volume} {29}} (\bibinfo {year} {2016})}\BibitemShut {NoStop}%
\bibitem [{\citenamefont {Alsing}\ \emph {et~al.}(2019)\citenamefont {Alsing}, \citenamefont {Charnock}, \citenamefont {Feeney},\ and\ \citenamefont {Wandelt}}]{alsing2019fast}%
  \BibitemOpen
  \bibfield  {author} {\bibinfo {author} {\bibfnamefont {J.}~\bibnamefont {Alsing}}, \bibinfo {author} {\bibfnamefont {T.}~\bibnamefont {Charnock}}, \bibinfo {author} {\bibfnamefont {S.}~\bibnamefont {Feeney}},\ and\ \bibinfo {author} {\bibfnamefont {B.}~\bibnamefont {Wandelt}},\ }\bibfield  {title} {\bibinfo {title} {Fast likelihood-free cosmology with neural density estimators and active learning},\ }\href@noop {} {\bibfield  {journal} {\bibinfo  {journal} {Monthly Notices of the Royal Astronomical Society}\ }\textbf {\bibinfo {volume} {488}},\ \bibinfo {pages} {4440} (\bibinfo {year} {2019})}\BibitemShut {NoStop}%
\bibitem [{\citenamefont {Cole}\ \emph {et~al.}(2022)\citenamefont {Cole}, \citenamefont {Miller}, \citenamefont {Witte}, \citenamefont {Cai}, \citenamefont {Grootes}, \citenamefont {Nattino},\ and\ \citenamefont {Weniger}}]{cole2022fast}%
  \BibitemOpen
  \bibfield  {author} {\bibinfo {author} {\bibfnamefont {A.}~\bibnamefont {Cole}}, \bibinfo {author} {\bibfnamefont {B.~K.}\ \bibnamefont {Miller}}, \bibinfo {author} {\bibfnamefont {S.~J.}\ \bibnamefont {Witte}}, \bibinfo {author} {\bibfnamefont {M.~X.}\ \bibnamefont {Cai}}, \bibinfo {author} {\bibfnamefont {M.~W.}\ \bibnamefont {Grootes}}, \bibinfo {author} {\bibfnamefont {F.}~\bibnamefont {Nattino}},\ and\ \bibinfo {author} {\bibfnamefont {C.}~\bibnamefont {Weniger}},\ }\bibfield  {title} {\bibinfo {title} {Fast and credible likelihood-free cosmology with truncated marginal neural ratio estimation},\ }\href@noop {} {\bibfield  {journal} {\bibinfo  {journal} {Journal of Cosmology and Astroparticle Physics}\ }\textbf {\bibinfo {volume} {2022}}\bibinfo  {number} { (09)},\ \bibinfo {pages} {004}}\BibitemShut {NoStop}%
\bibitem [{\citenamefont {Lemos}\ \emph {et~al.}(2023)\citenamefont {Lemos}, \citenamefont {Cranmer}, \citenamefont {Abidi}, \citenamefont {Hahn}, \citenamefont {Eickenberg}, \citenamefont {Massara}, \citenamefont {Yallup},\ and\ \citenamefont {Ho}}]{lemos2023robust}%
  \BibitemOpen
\bibfield  {number} {  }\bibfield  {author} {\bibinfo {author} {\bibfnamefont {P.}~\bibnamefont {Lemos}}, \bibinfo {author} {\bibfnamefont {M.}~\bibnamefont {Cranmer}}, \bibinfo {author} {\bibfnamefont {M.}~\bibnamefont {Abidi}}, \bibinfo {author} {\bibfnamefont {C.}~\bibnamefont {Hahn}}, \bibinfo {author} {\bibfnamefont {M.}~\bibnamefont {Eickenberg}}, \bibinfo {author} {\bibfnamefont {E.}~\bibnamefont {Massara}}, \bibinfo {author} {\bibfnamefont {D.}~\bibnamefont {Yallup}},\ and\ \bibinfo {author} {\bibfnamefont {S.}~\bibnamefont {Ho}},\ }\bibfield  {title} {\bibinfo {title} {Robust simulation-based inference in cosmology with bayesian neural networks},\ }\href@noop {} {\bibfield  {journal} {\bibinfo  {journal} {Machine Learning: Science and Technology}\ }\textbf {\bibinfo {volume} {4}},\ \bibinfo {pages} {01LT01} (\bibinfo {year} {2023})}\BibitemShut {NoStop}%
\bibitem [{\citenamefont {Papamakarios}(2019)}]{papamakarios2019neural}%
  \BibitemOpen
  \bibfield  {author} {\bibinfo {author} {\bibfnamefont {G.}~\bibnamefont {Papamakarios}},\ }\bibfield  {title} {\bibinfo {title} {Neural density estimation and likelihood-free inference},\ }\href@noop {} {\bibfield  {journal} {\bibinfo  {journal} {arXiv preprint arXiv:1910.13233}\ } (\bibinfo {year} {2019})}\BibitemShut {NoStop}%
\bibitem [{\citenamefont {Dupourqu{\'e}}\ \emph {et~al.}(2023)\citenamefont {Dupourqu{\'e}}, \citenamefont {Clerc}, \citenamefont {Pointecouteau}, \citenamefont {Eckert}, \citenamefont {Ettori},\ and\ \citenamefont {Vazza}}]{dupourque2023investigating}%
  \BibitemOpen
  \bibfield  {author} {\bibinfo {author} {\bibfnamefont {S.}~\bibnamefont {Dupourqu{\'e}}}, \bibinfo {author} {\bibfnamefont {N.}~\bibnamefont {Clerc}}, \bibinfo {author} {\bibfnamefont {E.}~\bibnamefont {Pointecouteau}}, \bibinfo {author} {\bibfnamefont {D.}~\bibnamefont {Eckert}}, \bibinfo {author} {\bibfnamefont {S.}~\bibnamefont {Ettori}},\ and\ \bibinfo {author} {\bibfnamefont {F.}~\bibnamefont {Vazza}},\ }\bibfield  {title} {\bibinfo {title} {Investigating the turbulent hot gas in x-cop galaxy clusters},\ }\href@noop {} {\bibfield  {journal} {\bibinfo  {journal} {Astronomy \& Astrophysics}\ }\textbf {\bibinfo {volume} {673}},\ \bibinfo {pages} {A91} (\bibinfo {year} {2023})}\BibitemShut {NoStop}%
\bibitem [{\citenamefont {Gatti}\ \emph {et~al.}(2024)\citenamefont {Gatti}, \citenamefont {Jeffrey}, \citenamefont {Whiteway}, \citenamefont {Williamson}, \citenamefont {Jain}, \citenamefont {Ajani}, \citenamefont {Anbajagane}, \citenamefont {Giannini}, \citenamefont {Zhou}, \citenamefont {Porredon} \emph {et~al.}}]{gatti2024dark}%
  \BibitemOpen
  \bibfield  {author} {\bibinfo {author} {\bibfnamefont {M.}~\bibnamefont {Gatti}}, \bibinfo {author} {\bibfnamefont {N.}~\bibnamefont {Jeffrey}}, \bibinfo {author} {\bibfnamefont {L.}~\bibnamefont {Whiteway}}, \bibinfo {author} {\bibfnamefont {J.}~\bibnamefont {Williamson}}, \bibinfo {author} {\bibfnamefont {B.}~\bibnamefont {Jain}}, \bibinfo {author} {\bibfnamefont {V.}~\bibnamefont {Ajani}}, \bibinfo {author} {\bibfnamefont {D.}~\bibnamefont {Anbajagane}}, \bibinfo {author} {\bibfnamefont {G.}~\bibnamefont {Giannini}}, \bibinfo {author} {\bibfnamefont {C.}~\bibnamefont {Zhou}}, \bibinfo {author} {\bibfnamefont {A.}~\bibnamefont {Porredon}}, \emph {et~al.},\ }\bibfield  {title} {\bibinfo {title} {Dark energy survey year 3 results: Simulation-based cosmological inference with wavelet harmonics, scattering transforms, and moments of weak lensing mass maps. validation on simulations},\ }\href@noop {} {\bibfield  {journal} {\bibinfo  {journal} {Physical Review D}\ }\textbf {\bibinfo {volume} {109}},\ \bibinfo
  {pages} {063534} (\bibinfo {year} {2024})}\BibitemShut {NoStop}%
\bibitem [{\citenamefont {Crisostomi}\ \emph {et~al.}(2023)\citenamefont {Crisostomi}, \citenamefont {Dey}, \citenamefont {Barausse},\ and\ \citenamefont {Trotta}}]{crisostomi2023neural}%
  \BibitemOpen
  \bibfield  {author} {\bibinfo {author} {\bibfnamefont {M.}~\bibnamefont {Crisostomi}}, \bibinfo {author} {\bibfnamefont {K.}~\bibnamefont {Dey}}, \bibinfo {author} {\bibfnamefont {E.}~\bibnamefont {Barausse}},\ and\ \bibinfo {author} {\bibfnamefont {R.}~\bibnamefont {Trotta}},\ }\bibfield  {title} {\bibinfo {title} {Neural posterior estimation with guaranteed exact coverage: The ringdown of gw150914},\ }\href@noop {} {\bibfield  {journal} {\bibinfo  {journal} {Physical Review D}\ }\textbf {\bibinfo {volume} {108}},\ \bibinfo {pages} {044029} (\bibinfo {year} {2023})}\BibitemShut {NoStop}%
\bibitem [{\citenamefont {Christy}\ \emph {et~al.}(2024)\citenamefont {Christy}, \citenamefont {Baxter},\ and\ \citenamefont {Kumar}}]{christy2024applying}%
  \BibitemOpen
  \bibfield  {author} {\bibinfo {author} {\bibfnamefont {K.}~\bibnamefont {Christy}}, \bibinfo {author} {\bibfnamefont {E.~J.}\ \bibnamefont {Baxter}},\ and\ \bibinfo {author} {\bibfnamefont {J.}~\bibnamefont {Kumar}},\ }\bibfield  {title} {\bibinfo {title} {Applying simulation-based inference to spectral and spatial information from the galactic center gamma-ray excess},\ }\href@noop {} {\bibfield  {journal} {\bibinfo  {journal} {arXiv preprint arXiv:2402.04549}\ } (\bibinfo {year} {2024})}\BibitemShut {NoStop}%
\bibitem [{\citenamefont {Harnois-Deraps}\ \emph {et~al.}(2024)\citenamefont {Harnois-Deraps}, \citenamefont {Heydenreich}, \citenamefont {Giblin}, \citenamefont {Martinet}, \citenamefont {Troester}, \citenamefont {Asgari}, \citenamefont {Burger}, \citenamefont {Castro}, \citenamefont {Dolag}, \citenamefont {Heymans} \emph {et~al.}}]{harnois2024kids}%
  \BibitemOpen
  \bibfield  {author} {\bibinfo {author} {\bibfnamefont {J.}~\bibnamefont {Harnois-Deraps}}, \bibinfo {author} {\bibfnamefont {S.}~\bibnamefont {Heydenreich}}, \bibinfo {author} {\bibfnamefont {B.}~\bibnamefont {Giblin}}, \bibinfo {author} {\bibfnamefont {N.}~\bibnamefont {Martinet}}, \bibinfo {author} {\bibfnamefont {T.}~\bibnamefont {Troester}}, \bibinfo {author} {\bibfnamefont {M.}~\bibnamefont {Asgari}}, \bibinfo {author} {\bibfnamefont {P.}~\bibnamefont {Burger}}, \bibinfo {author} {\bibfnamefont {T.}~\bibnamefont {Castro}}, \bibinfo {author} {\bibfnamefont {K.}~\bibnamefont {Dolag}}, \bibinfo {author} {\bibfnamefont {C.}~\bibnamefont {Heymans}}, \emph {et~al.},\ }\bibfield  {title} {\bibinfo {title} {Kids-1000 and des-y1 combined: Cosmology from peak count statistics},\ }\href@noop {} {\bibfield  {journal} {\bibinfo  {journal} {arXiv preprint arXiv:2405.10312}\ } (\bibinfo {year} {2024})}\BibitemShut {NoStop}%
\bibitem [{\citenamefont {Moser}\ \emph {et~al.}(2024)\citenamefont {Moser}, \citenamefont {Kacprzak}, \citenamefont {Fischbacher}, \citenamefont {Refregier}, \citenamefont {Grimm},\ and\ \citenamefont {Tortorelli}}]{moser2024simulation}%
  \BibitemOpen
  \bibfield  {author} {\bibinfo {author} {\bibfnamefont {B.}~\bibnamefont {Moser}}, \bibinfo {author} {\bibfnamefont {T.}~\bibnamefont {Kacprzak}}, \bibinfo {author} {\bibfnamefont {S.}~\bibnamefont {Fischbacher}}, \bibinfo {author} {\bibfnamefont {A.}~\bibnamefont {Refregier}}, \bibinfo {author} {\bibfnamefont {D.}~\bibnamefont {Grimm}},\ and\ \bibinfo {author} {\bibfnamefont {L.}~\bibnamefont {Tortorelli}},\ }\bibfield  {title} {\bibinfo {title} {Simulation-based inference of deep fields: galaxy population model and redshift distributions},\ }\href@noop {} {\bibfield  {journal} {\bibinfo  {journal} {Journal of Cosmology and Astroparticle Physics}\ }\textbf {\bibinfo {volume} {2024}}\bibinfo  {number} { (05)},\ \bibinfo {pages} {049}}\BibitemShut {NoStop}%
\bibitem [{\citenamefont {Novaes}\ \emph {et~al.}(2024)\citenamefont {Novaes}, \citenamefont {Thiele}, \citenamefont {Armijo}, \citenamefont {Cheng}, \citenamefont {Cowell}, \citenamefont {Marques}, \citenamefont {Ferreira}, \citenamefont {Shirasaki}, \citenamefont {Osato},\ and\ \citenamefont {Liu}}]{novaes2024cosmology}%
  \BibitemOpen
\bibfield  {number} {  }\bibfield  {author} {\bibinfo {author} {\bibfnamefont {C.~P.}\ \bibnamefont {Novaes}}, \bibinfo {author} {\bibfnamefont {L.}~\bibnamefont {Thiele}}, \bibinfo {author} {\bibfnamefont {J.}~\bibnamefont {Armijo}}, \bibinfo {author} {\bibfnamefont {S.}~\bibnamefont {Cheng}}, \bibinfo {author} {\bibfnamefont {J.~A.}\ \bibnamefont {Cowell}}, \bibinfo {author} {\bibfnamefont {G.~A.}\ \bibnamefont {Marques}}, \bibinfo {author} {\bibfnamefont {E.~G.}\ \bibnamefont {Ferreira}}, \bibinfo {author} {\bibfnamefont {M.}~\bibnamefont {Shirasaki}}, \bibinfo {author} {\bibfnamefont {K.}~\bibnamefont {Osato}},\ and\ \bibinfo {author} {\bibfnamefont {J.}~\bibnamefont {Liu}},\ }\bibfield  {title} {\bibinfo {title} {Cosmology from hsc y1 weak lensing with combined higher-order statistics and simulation-based inference},\ }\href@noop {} {\bibfield  {journal} {\bibinfo  {journal} {arXiv preprint arXiv:2409.01301}\ } (\bibinfo {year} {2024})}\BibitemShut {NoStop}%
\bibitem [{\citenamefont {Fischbacher}\ \emph {et~al.}(2024)\citenamefont {Fischbacher}, \citenamefont {Moser}, \citenamefont {Kacprzak}, \citenamefont {Herbel}, \citenamefont {Tortorelli}, \citenamefont {Schmitt}, \citenamefont {Refregier},\ and\ \citenamefont {Amara}}]{fischbacher2024texttt}%
  \BibitemOpen
  \bibfield  {author} {\bibinfo {author} {\bibfnamefont {S.}~\bibnamefont {Fischbacher}}, \bibinfo {author} {\bibfnamefont {B.}~\bibnamefont {Moser}}, \bibinfo {author} {\bibfnamefont {T.}~\bibnamefont {Kacprzak}}, \bibinfo {author} {\bibfnamefont {J.}~\bibnamefont {Herbel}}, \bibinfo {author} {\bibfnamefont {L.}~\bibnamefont {Tortorelli}}, \bibinfo {author} {\bibfnamefont {U.}~\bibnamefont {Schmitt}}, \bibinfo {author} {\bibfnamefont {A.}~\bibnamefont {Refregier}},\ and\ \bibinfo {author} {\bibfnamefont {A.}~\bibnamefont {Amara}},\ }\bibfield  {title} {\bibinfo {title} {$\texttt{galsbi}$ : A python package for the galsbi galaxy population model},\ }\href@noop {} {\bibfield  {journal} {\bibinfo  {journal} {arXiv preprint arXiv:2412.08722}\ } (\bibinfo {year} {2024})}\BibitemShut {NoStop}%
\bibitem [{\citenamefont {Castelvecchi}(2016)}]{castelvecchi2016can}%
  \BibitemOpen
  \bibfield  {author} {\bibinfo {author} {\bibfnamefont {D.}~\bibnamefont {Castelvecchi}},\ }\bibfield  {title} {\bibinfo {title} {Can we open the black box of ai?},\ }\href@noop {} {\bibfield  {journal} {\bibinfo  {journal} {Nature News}\ }\textbf {\bibinfo {volume} {538}},\ \bibinfo {pages} {20} (\bibinfo {year} {2016})}\BibitemShut {NoStop}%
\bibitem [{\citenamefont {Hermans}\ \emph {et~al.}(2021)\citenamefont {Hermans}, \citenamefont {Delaunoy}, \citenamefont {Rozet}, \citenamefont {Wehenkel}, \citenamefont {Begy},\ and\ \citenamefont {Louppe}}]{hermans2021trust}%
  \BibitemOpen
  \bibfield  {author} {\bibinfo {author} {\bibfnamefont {J.}~\bibnamefont {Hermans}}, \bibinfo {author} {\bibfnamefont {A.}~\bibnamefont {Delaunoy}}, \bibinfo {author} {\bibfnamefont {F.}~\bibnamefont {Rozet}}, \bibinfo {author} {\bibfnamefont {A.}~\bibnamefont {Wehenkel}}, \bibinfo {author} {\bibfnamefont {V.}~\bibnamefont {Begy}},\ and\ \bibinfo {author} {\bibfnamefont {G.}~\bibnamefont {Louppe}},\ }\bibfield  {title} {\bibinfo {title} {A trust crisis in simulation-based inference? your posterior approximations can be unfaithful},\ }\href@noop {} {\bibfield  {journal} {\bibinfo  {journal} {arXiv preprint arXiv:2110.06581}\ } (\bibinfo {year} {2021})}\BibitemShut {NoStop}%
\bibitem [{\citenamefont {Leclercq}\ \emph {et~al.}(2019)\citenamefont {Leclercq}, \citenamefont {Enzi}, \citenamefont {Jasche},\ and\ \citenamefont {Heavens}}]{leclercq2019primordial}%
  \BibitemOpen
  \bibfield  {author} {\bibinfo {author} {\bibfnamefont {F.}~\bibnamefont {Leclercq}}, \bibinfo {author} {\bibfnamefont {W.}~\bibnamefont {Enzi}}, \bibinfo {author} {\bibfnamefont {J.}~\bibnamefont {Jasche}},\ and\ \bibinfo {author} {\bibfnamefont {A.}~\bibnamefont {Heavens}},\ }\bibfield  {title} {\bibinfo {title} {Primordial power spectrum and cosmology from black-box galaxy surveys},\ }\href@noop {} {\bibfield  {journal} {\bibinfo  {journal} {Monthly Notices of the Royal Astronomical Society}\ }\textbf {\bibinfo {volume} {490}},\ \bibinfo {pages} {4237} (\bibinfo {year} {2019})}\BibitemShut {NoStop}%
\bibitem [{\citenamefont {Heavens}\ \emph {et~al.}(2000)\citenamefont {Heavens}, \citenamefont {Jimenez},\ and\ \citenamefont {Lahav}}]{heavens2000massive}%
  \BibitemOpen
  \bibfield  {author} {\bibinfo {author} {\bibfnamefont {A.~F.}\ \bibnamefont {Heavens}}, \bibinfo {author} {\bibfnamefont {R.}~\bibnamefont {Jimenez}},\ and\ \bibinfo {author} {\bibfnamefont {O.}~\bibnamefont {Lahav}},\ }\bibfield  {title} {\bibinfo {title} {Massive lossless data compression and multiple parameter estimation from galaxy spectra},\ }\href@noop {} {\bibfield  {journal} {\bibinfo  {journal} {Monthly Notices of the Royal Astronomical Society}\ }\textbf {\bibinfo {volume} {317}},\ \bibinfo {pages} {965} (\bibinfo {year} {2000})}\BibitemShut {NoStop}%
\bibitem [{\citenamefont {Gupta}\ and\ \citenamefont {Nagar}(2018)}]{gupta2018matrix}%
  \BibitemOpen
  \bibfield  {author} {\bibinfo {author} {\bibfnamefont {A.~K.}\ \bibnamefont {Gupta}}\ and\ \bibinfo {author} {\bibfnamefont {D.~K.}\ \bibnamefont {Nagar}},\ }\href@noop {} {\emph {\bibinfo {title} {Matrix variate distributions}}}\ (\bibinfo  {publisher} {Chapman and Hall/CRC},\ \bibinfo {year} {2018})\BibitemShut {NoStop}%
\bibitem [{\citenamefont {Piras}\ and\ \citenamefont {Mancini}(2023)}]{piras2023cosmopower}%
  \BibitemOpen
  \bibfield  {author} {\bibinfo {author} {\bibfnamefont {D.}~\bibnamefont {Piras}}\ and\ \bibinfo {author} {\bibfnamefont {A.~S.}\ \bibnamefont {Mancini}},\ }\bibfield  {title} {\bibinfo {title} {Cosmopower-jax: high-dimensional bayesian inference with differentiable cosmological emulators},\ }\href@noop {} {\bibfield  {journal} {\bibinfo  {journal} {arXiv preprint arXiv:2305.06347}\ } (\bibinfo {year} {2023})}\BibitemShut {NoStop}%
\bibitem [{\citenamefont {Spurio~Mancini}\ \emph {et~al.}(2022)\citenamefont {Spurio~Mancini}, \citenamefont {Piras}, \citenamefont {Alsing}, \citenamefont {Joachimi},\ and\ \citenamefont {Hobson}}]{spurio2022cosmopower}%
  \BibitemOpen
  \bibfield  {author} {\bibinfo {author} {\bibfnamefont {A.}~\bibnamefont {Spurio~Mancini}}, \bibinfo {author} {\bibfnamefont {D.}~\bibnamefont {Piras}}, \bibinfo {author} {\bibfnamefont {J.}~\bibnamefont {Alsing}}, \bibinfo {author} {\bibfnamefont {B.}~\bibnamefont {Joachimi}},\ and\ \bibinfo {author} {\bibfnamefont {M.~P.}\ \bibnamefont {Hobson}},\ }\bibfield  {title} {\bibinfo {title} {Cosmopower: emulating cosmological power spectra for accelerated bayesian inference from next-generation surveys},\ }\href@noop {} {\bibfield  {journal} {\bibinfo  {journal} {Monthly Notices of the Royal Astronomical Society}\ }\textbf {\bibinfo {volume} {511}},\ \bibinfo {pages} {1771} (\bibinfo {year} {2022})}\BibitemShut {NoStop}%
\bibitem [{\citenamefont {Speagle}(2020)}]{speagle2020dynesty}%
  \BibitemOpen
  \bibfield  {author} {\bibinfo {author} {\bibfnamefont {J.~S.}\ \bibnamefont {Speagle}},\ }\bibfield  {title} {\bibinfo {title} {dynesty: a dynamic nested sampling package for estimating bayesian posteriors and evidences},\ }\href@noop {} {\bibfield  {journal} {\bibinfo  {journal} {Monthly Notices of the Royal Astronomical Society}\ }\textbf {\bibinfo {volume} {493}},\ \bibinfo {pages} {3132} (\bibinfo {year} {2020})}\BibitemShut {NoStop}%
\bibitem [{\citenamefont {Koposov}\ \emph {et~al.}(2022)\citenamefont {Koposov}, \citenamefont {Speagle}, \citenamefont {Barbary}, \citenamefont {Ashton}, \citenamefont {Bennett}, \citenamefont {Buchner}, \citenamefont {Scheffler}, \citenamefont {Cook}, \citenamefont {Talbot}, \citenamefont {Guillochon} \emph {et~al.}}]{koposov2022joshspeagle}%
  \BibitemOpen
  \bibfield  {author} {\bibinfo {author} {\bibfnamefont {S.}~\bibnamefont {Koposov}}, \bibinfo {author} {\bibfnamefont {J.}~\bibnamefont {Speagle}}, \bibinfo {author} {\bibfnamefont {K.}~\bibnamefont {Barbary}}, \bibinfo {author} {\bibfnamefont {G.}~\bibnamefont {Ashton}}, \bibinfo {author} {\bibfnamefont {E.}~\bibnamefont {Bennett}}, \bibinfo {author} {\bibfnamefont {J.}~\bibnamefont {Buchner}}, \bibinfo {author} {\bibfnamefont {C.}~\bibnamefont {Scheffler}}, \bibinfo {author} {\bibfnamefont {B.}~\bibnamefont {Cook}}, \bibinfo {author} {\bibfnamefont {C.}~\bibnamefont {Talbot}}, \bibinfo {author} {\bibfnamefont {J.}~\bibnamefont {Guillochon}}, \emph {et~al.},\ }\bibfield  {title} {\bibinfo {title} {joshspeagle/dynesty: v2. 0.0},\ }\href@noop {} {\bibfield  {journal} {\bibinfo  {journal} {Zenodo}\ } (\bibinfo {year} {2022})}\BibitemShut {NoStop}%
\bibitem [{\citenamefont {Higson}\ \emph {et~al.}(2019)\citenamefont {Higson}, \citenamefont {Handley}, \citenamefont {Hobson},\ and\ \citenamefont {Lasenby}}]{higson2019dynamic}%
  \BibitemOpen
  \bibfield  {author} {\bibinfo {author} {\bibfnamefont {E.}~\bibnamefont {Higson}}, \bibinfo {author} {\bibfnamefont {W.}~\bibnamefont {Handley}}, \bibinfo {author} {\bibfnamefont {M.}~\bibnamefont {Hobson}},\ and\ \bibinfo {author} {\bibfnamefont {A.}~\bibnamefont {Lasenby}},\ }\bibfield  {title} {\bibinfo {title} {Dynamic nested sampling: an improved algorithm for parameter estimation and evidence calculation},\ }\href@noop {} {\bibfield  {journal} {\bibinfo  {journal} {Statistics and Computing}\ }\textbf {\bibinfo {volume} {29}},\ \bibinfo {pages} {891} (\bibinfo {year} {2019})}\BibitemShut {NoStop}%
\bibitem [{\citenamefont {Lewis}(2019)}]{lewis2019getdist}%
  \BibitemOpen
  \bibfield  {author} {\bibinfo {author} {\bibfnamefont {A.}~\bibnamefont {Lewis}},\ }\bibfield  {title} {\bibinfo {title} {Getdist: a python package for analysing monte carlo samples},\ }\href@noop {} {\bibfield  {journal} {\bibinfo  {journal} {arXiv preprint arXiv:1910.13970}\ } (\bibinfo {year} {2019})}\BibitemShut {NoStop}%
\bibitem [{\citenamefont {Alsing}\ and\ \citenamefont {Wandelt}(2018)}]{alsing2018generalized}%
  \BibitemOpen
  \bibfield  {author} {\bibinfo {author} {\bibfnamefont {J.}~\bibnamefont {Alsing}}\ and\ \bibinfo {author} {\bibfnamefont {B.}~\bibnamefont {Wandelt}},\ }\bibfield  {title} {\bibinfo {title} {Generalized massive optimal data compression},\ }\href@noop {} {\bibfield  {journal} {\bibinfo  {journal} {Monthly Notices of the Royal Astronomical Society: Letters}\ }\textbf {\bibinfo {volume} {476}},\ \bibinfo {pages} {L60} (\bibinfo {year} {2018})}\BibitemShut {NoStop}%
\bibitem [{\citenamefont {Alsing}\ \emph {et~al.}(2018)\citenamefont {Alsing}, \citenamefont {Wandelt},\ and\ \citenamefont {Feeney}}]{alsing2018massive}%
  \BibitemOpen
  \bibfield  {author} {\bibinfo {author} {\bibfnamefont {J.}~\bibnamefont {Alsing}}, \bibinfo {author} {\bibfnamefont {B.}~\bibnamefont {Wandelt}},\ and\ \bibinfo {author} {\bibfnamefont {S.}~\bibnamefont {Feeney}},\ }\bibfield  {title} {\bibinfo {title} {Massive optimal data compression and density estimation for scalable, likelihood-free inference in cosmology},\ }\href@noop {} {\bibfield  {journal} {\bibinfo  {journal} {Monthly Notices of the Royal Astronomical Society}\ }\textbf {\bibinfo {volume} {477}},\ \bibinfo {pages} {2874} (\bibinfo {year} {2018})}\BibitemShut {NoStop}%
\end{thebibliography}%

\end{document}